\PassOptionsToPackage{capitalize,noabbrev,nameinlink}{cleveref}                                                                                                                                                 
\PassOptionsToPackage{usenames,dvipsnames}{color}
\PassOptionsToPackage{usenames,table}{xcolor}
\PassOptionsToPackage{final}{microtype}
\PassOptionsToPackage{scaled}{inconsolata}
\documentclass[10pt,sigconf, nonacm]{acmart}
\usepackage{amsmath,epsfig}

\newcommand{\paperTitle}{Mainlining Databases: Supporting
\texorpdfstring{\\}{}Fast Transactional Workloads on
\texorpdfstring{\\}{}Universal Columnar Data File Formats}
\newcommand{\paperKeywords}{Database Systems, Apache Arrow, Joy Smells}
\newcommand{\paperAuthors}{Tianyu Li, Matthew Butrovich, Amadou Ngom, Wes McKinney, Andrew Pavlo}

\setcopyright{acmcopyright}
\copyrightyear{2019}
\acmYear{2019}
\acmConference[SIGMOD'20]{SIGMOD'20: 2020 International Conference on
    Management of Data}{June 14--19, 2020}{Portland, OR, USA}
\acmBooktitle{SIGMOD'20: 2020 International Conference on Management of Data,
    June 14--19, 2020, Portland, OR, USA}
\acmPrice{0.00}
\acmISBN{978-1-4503-9999-9/18/06}

\hypersetup{
    pdfauthor = {\paperAuthors},
    pdftitle = {\paperTitle},
    pdfkeywords = {\paperKeywords},
    pdfborder={ 0 0 0 }
}

\usepackage{amsmath}
\usepackage{amssymb}
\usepackage{boxedminipage}
\usepackage{xspace}
\usepackage{tabularx}
\usepackage{balance}  % for  \balance command ON LAST PAGE  (only there!)
\usepackage{url}

\usepackage[hyphenbreaks]{breakurl}
\usepackage[font={small}]{caption}
\usepackage{graphicx}
\usepackage{subfig}
\usepackage[usenames,table]{xcolor}
\usepackage{cleveref}
\usepackage{tabularx}
\usepackage{epsfig}

\usepackage[final]{microtype}
\usepackage[T1]{fontenc}
\usepackage{graphicx}
\usepackage{subfig}
\usepackage{soul}
\usepackage{pifont}

\usepackage{booktabs}
\usepackage[end]{algpseudocode}
\usepackage{algorithm}

\newcommand{\mlbegin}{\shortstack\bgroup}
\newcommand{\mlend}{\egroup}

\newcounter{Lcount}
\newcommand{\squishlist}{
    \begin{list}{\arabic{Lcount}. }
   { \usecounter{Lcount}
        \setlength{\itemsep}{0pt}
        \setlength{\parsep}{3pt}
        \setlength{\topsep}{3pt}
        \setlength{\partopsep}{0pt}
        \setlength{\leftmargin}{2em}
        \setlength{\labelwidth}{1.5em}
        \setlength{\labelsep}{0.5em} } }

\newcommand{\squishend}{\end{list}}

\definecolor{todo-color}{rgb}{1,0,0}

\definecolor{comment-color}{rgb}{0.25,0.25,0.25}

\newcommand{\dbTable}[1]{\texttt{\MakeUppercase{#1}}\xspace}

\newcommand{\dbTxnManager}{transaction engine\xspace}
\newcommand{\dbLogManager}{log manager\xspace}
\newcommand{\dbTupleSlot}{\texttt{TupleSlot}\xspace}
\newcommand{\dbVarlenEntry}{\texttt{VarlenEntry}\xspace}

\newcommand{\xstatusFont}[1]{{\small \textsf{#1}}}
\newcommand{\statusHot}{\xstatusFont{hot}\xspace}
\newcommand{\statusCooling}{\xstatusFont{cooling}\xspace}
\newcommand{\statusFreezing}{\xstatusFont{freezing}\xspace}
\newcommand{\statusFrozen}{\xstatusFont{frozen}\xspace}

\newcommand{\transformHybridGather}{\xstatusFont{Hybrid-Gather}\xspace}
\newcommand{\transformHybridCompress}{\xstatusFont{Hybrid-Compress}\xspace}
\newcommand{\transformSnapshot}{\xstatusFont{Snapshot}\xspace}
\newcommand{\transformTxnInplace}{\xstatusFont{In-Place}\xspace}

\newcommand{\sysname}{DB-X\xspace}
\newcommand{\postgres}{PostgreSQL\xspace}
\newcommand{\htwoo}{H2O\xspace}
\newcommand{\peloton}{Peloton\xspace}

\newcommand{\tsStart}{\textit{start}\xspace}
\newcommand{\tsCommit}{\textit{commit}\xspace}

\begin{document}

\newcommand{\mail}[1]{\href{mailto:#1}{#1}}
\newcommand{\superscript}[1]{\ensuremath{^{\textrm{#1}}}}

\def\affilMIT{\superscript{$\clubsuit$}}
\def\affilURSA{\superscript{$\spadesuit$}}

% \numberofauthors{1}
% \author{
%     \alignauthor Tianyu Li\affilMIT, Matthew Butrovich, Amadou Ngom, Wan Shen Lim \\
%     Wes McKinney\affilURSA, Andrew Pavlo \\
%     \affaddr{Massachusetts Institute of Technology\affilMIT,
%              Carnegie Mellon University,
%              Ursa Labs\affilURSA}\\
%     \vspace*{-0.15em}   
%     \email{\eaddfnt{
%             \mail{litianyu@mit.edu}\\
%           \{\href{mailto:mbutrovi@cs.cmu.edu}{mbutrovi},
%             \href{mailto:angom@cs.cmu.edu}{angom},
%             \href{mailto:wanshenl@cs.cmu.edu}{wanshenl},
%             \href{mailto:pavlo@cs.cmu.edu}{pavlo}\}@cs.cmu.edu \\
%             \mail{wes@ursalabs.org}
%     }}
% }

\author{Tianyu Li}
\email{litianyu@mit.edu}
\affiliation{%
  \institution{MIT}
}

\author{Matthew Butrovich}
\email{mbutrovi@cs.cmu.edu}
\affiliation{
  \institution{Carnegie Mellon University}
}

\author{Amadou Ngom}
\email{angom@cs.cmu.edu}
\affiliation{
  \institution{Carnegie Mellon University}
}

\author{Wan Shen Lim}
\email{wanshenl@cs.cmu.edu}
\affiliation{
  \institution{Carnegie Mellon University}
}

\author{Wes McKinney}
\email{wes@ursalabs.org}
\affiliation{
  \institution{Ursa Labs}
}

\author{Andrew Pavlo}
\email{pavlo@cs.cmu.edu}
\affiliation{
  \institution{Carnegie Mellon University}
}

\renewcommand{\shortauthors}{T. Li et al.}

% [Short Title]{Full Title}
\title[Mainlining Databases: Supporting Fast Transactional Workloads on\\Universal
       Columnar Data File Formats]{\paperTitle}

%% ==================================================================
%% ABSTRACT
%% ==================================================================
\begin{abstract}
The proliferation of modern data processing tools has given rise to open-source columnar
data formats. The advantage of these formats is that they help organizations avoid
repeatedly converting data to a new format for each application. These formats, however,
are read-only, and organizations must use a heavy-weight transformation process to load
data from on-line transactional processing (OLTP) systems. We aim to reduce or even
eliminate this process by developing a storage architecture for in-memory database
management systems (DBMSs) that is aware of the eventual usage of its data and emits
columnar storage blocks in a universal open-source format. We introduce relaxations to
common analytical data formats to efficiently update records and rely on a lightweight
transformation process to convert blocks to a read-optimized layout when they are cold.
We also describe how to access data from third-party analytical tools with minimal
serialization overhead. To evaluate our work, we implemented our storage engine based on
the Apache Arrow format and integrated it into the \sysname DBMS. Our experiments show
that our approach achieves comparable performance with dedicated OLTP DBMSs while enabling
orders-of-magnitude faster data exports to external data science and machine learning
tools than existing methods.
\end{abstract}

\maketitle

%% ==================================================================
%% INTRODUCTION
%% ==================================================================
\section{Introduction}
\label{sec:introduction}
Data analysis pipelines allow organizations to extrapolate insights from data residing in
their OLTP systems. The tools in these pipelines often use open-source binary formats,
such as Apache Parquet~\cite{parquet}, Apache ORC~\cite{apacheorc} and Apache
Arrow~\cite{apachearrow}. Such formats allow disparate systems to exchange data through a
common interface without converting between proprietary formats. But these formats target
write-once, read-many workloads and are not amenable to OLTP systems. This means that a
data scientist must transform OLTP data with a heavy-weight process, which is
computationally expensive and inhibits analytical operations. 

% Over the past decade, several companies and research groups have
% developed hybrid transactional analytical processing (HTAP) DBMSs in attempts to
% address this issue~\cite{pezzini14}. These systems, however, are not one-size-fit-all
% solutions to the problem.
% This problem is likely to persist.
Although a DBMS can perform some analytical duties, modern data science workloads often
involve specialized frameworks, such as TensorFlow, PyTorch, and Pandas. Organizations are
also heavily invested in personnel, tooling, and infrastructure for the current data
science eco-system of Python tools. We contend that the need for DBMS to efficiently
export large amounts of data to external tools will persist. To enable analysis of data as
soon as it arrives in a database is, and to deliver performance gains across the entire
data analysis pipeline, we should look to improve a DBMS's interoperability with external
tools.

If an OLTP DBMS directly stores data in a format used by downstream applications, the
export cost is just the cost of network transmission. The challenge in this is that most
open-source formats are optimized for read/append operations, not in-place updates.
Meanwhile, divergence from the target format in the OLTP DBMS translates into more
transformation overhead when exporting data, which can be equally detrimental to
performance. A viable design must seek equilibrium in these two conflicting
considerations.

To address this challenge, we present a multi-versioned DBMS that operates on a relaxation
of an open-source columnar format to support efficient OLTP  modifications. The relaxed
format can then be transformed into the canonical format as data cools with a light-weight
in-memory process. We implemented our storage and concurrency control architecture in 
\textbf{\sysname}~\cite{cmu-dbms} and evaluated its performance. We target Apache Arrow,
although our approach is also applicable to other columnar formats. Our results show that
we achieve good performance on OLTP workloads operating on the relaxed Arrow format. We
also implemented an Arrow export layer for our system, and show that it facilitates
orders-of-magnitude faster exports to external tools. %compared to existing systems.

The remainder of this paper is organized as follows: we first discuss in \cref{sec:bg} 
the motivation for this work. % and introduce the Arrow storage format.
We then present our storage architecture and concurrency control in \cref{sec:sys},
followed by our transformation algorithm in \cref{sec:transformation}. In
\cref{sec:export}, we discuss how to export data to external tools. We present our
experimental evaluation in \cref{sec:eval} and discuss related work in
\cref{sec:relatedwork}.

%% ==================================================================
%% BACKGROUND
%% ==================================================================
\section{Background}
\label{sec:bg}
We now discuss challenges in running analysis with external tools with OLTP DBMSs. We
begin by describing how data transformation and movement are bottlenecks in data
processing. We then present a popular open-source format (Apache Arrow) and discuss its
strengths and weaknesses.

%% ---------------------------------------------------------------
%% Data Transformation Overheads
%% ---------------------------------------------------------------
\subsection{Data Movement and Transformation}
\label{sec:bg-export}
A data processing pipeline consists of a front-end OLTP layer and multiple analytical
layers. OLTP engines employ the $n$-ary storage model (i.e., row-store) to support
efficient single-tuple operations, while the analytical layers use the decomposition
storage model (i.e., column-store) to speed up large
scans~\cite{abadi08,boncz05,menon17,kersten18}. Because of conflicting optimization
strategies for these two use cases, organizations often implement the pipeline by
combining specialized systems.

The most salient issue with this bifurcated approach is data transformation and movement
between layers. This problem is made worse with the emergence of machine learning
workloads that load the entire data set instead of a small query result set. For example,
a data scientist will (1) execute SQL queries to export data from \postgres,  (2) load it
into a Jupyter notebook on a local machine and prepare it with Pandas, and (3) train
models on cleaned data with TensorFlow. Each step in such a pipeline transforms data into
a format native to the target framework: a disk-optimized row-store for \postgres,
DataFrames for Pandas, and tensors for TensorFlow. The slowest of all transformations is
from the DBMS to Pandas because it retrieves data over the DBMS's network protocol and
then rewrites it into the desired columnar format. This process is not optimal for
high-bandwidth data movement~\cite{raasveldt17}. Many organizations employ costly
extract-transform-load (ETL) pipelines that run only nightly, introducing delays to
analytics.  

To better understand this issue, we measured the time it takes to extract data from
\postgres (v10.6) and load it into a Pandas program. We use the \dbTable{LINEITEM} table
from TPC-H with scale factor 10 (60M tuples, 8~GB as a CSV file, 11~GB as a \postgres
table). We compare three approaches for loading the table into the Python program: (1) SQL
over a Python ODBC connection, (2)  using \postgres's \texttt{COPY} command to export a
CSV file to disk and then loading it into Pandas, and (3) loading data directly from a
buffer already in the Python runtime's memory. The last method represents the theoretical
best-case scenario to provide us with an upper bound for data export speed. We pre-load
the entire table into \postgres's buffer pool using the \texttt{pg\_warm} extension. To
simplify our setup, we run the Python program on the same machine as the DBMS. We use a
machine with 128~GB of memory, of which we reserve 15~GB for \postgres's shared buffers.
We provide a full description of our operating environment for this experiment in
\cref{sec:eval}.

\begin{figure}[t!]
    \centering
    \includegraphics[width=\columnwidth]{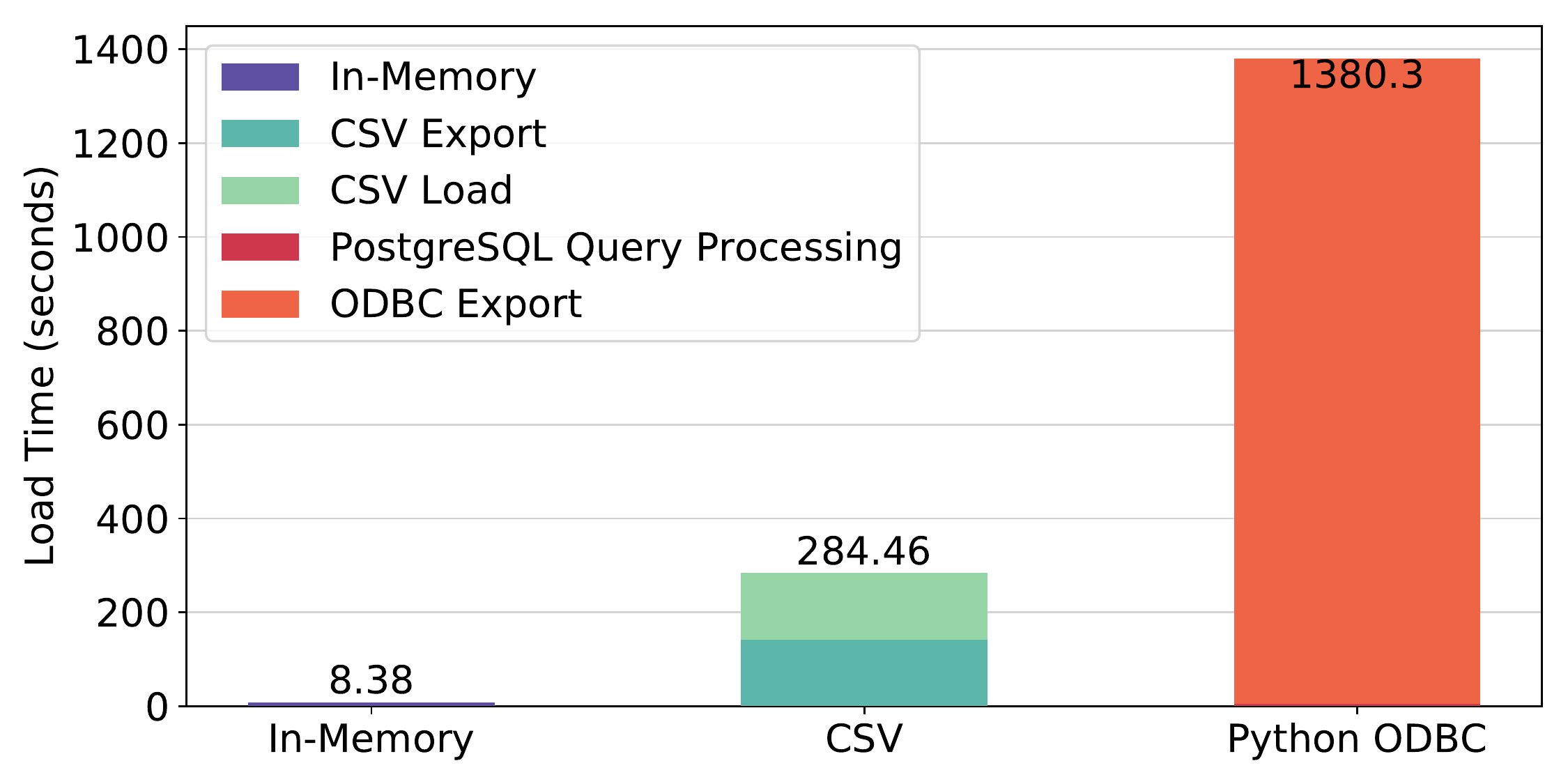}
    \caption{
        \textbf{Data Transformation Costs} -- 
        Time taken to load a TPC-H table into Pandas with different approaches.
    }
    \label{fig:motivation}
\end{figure}

The results in \cref{fig:motivation} show that ODBC and CSV are orders of magnitude slower
than what is possible. This difference is because of the overhead of transforming to a
different format, as well as excessive serialization in the \postgres wire protocol. Query
processing itself takes $0.004\%$ of the total export time. The rest of the time is spent
in the serialization layer and in transforming the data. Optimizing this export process
will significantly speed up analytics pipelines.

%% ---------------------------------------------------------------
%% Apache Arrow
%% ---------------------------------------------------------------
\subsection{Column-Stores and Apache Arrow}
\label{sec:bg-arrow}
The inefficiency of loading data through a SQL interface requires us to rethink the data
export process and avoid costly data transformations. Lack of interoperability between
row-stores and analytical columnar formats is a major source of inefficiency. As discussed
previously, OLTP DBMSs are row-stores because conventional wisdom is that column-stores
are inferior for OLTP workloads. Recent work, however, has shown that column-stores can
also support high-performance transactional processing~\cite{neumann15, sikka12}. We
propose implementing a high-performance OLTP DBMS directly on top of a data format used by
analytics tools. To do so, we select a representative format (Apache Arrow) and analyze
its strengths and weaknesses for OLTP workloads.

\begin{figure}[t!]
    \centering
    \includegraphics[width=0.9\columnwidth]{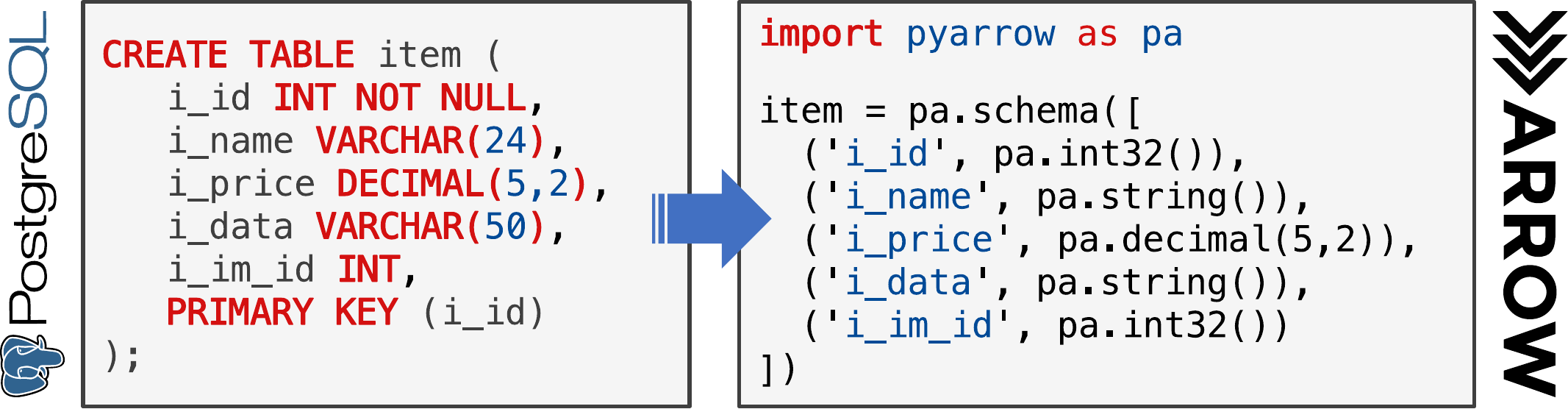}
    \caption{
        \textbf{SQL Table to Arrow} -- 
        An example of using Arrow's API to describe a SQL table's
        schema in Python.
    }
    \label{fig:arrow-ddl}
\end{figure}

Apache Arrow is a cross-language development platform for in-memory
data~\cite{apachearrow}. In the early 2010s, developers from Apache Drill, Apache Impala,
Apache Kudu, Pandas, and others independently explored universal in-memory columnar data
formats. These groups joined together in 2015 to develop a shared format based on their
overlapping requirements. Arrow was introduced in 2016 and has since become the standard
for columnar in-memory analytics, and as a high-performance interface between
heterogeneous systems. There is a growing ecosystem of tools built for Arrow, including
APIs for several programming languages and computational libraries. For example, Google's
TensorFlow now integrates with Arrow through a Python module~\cite{tf-arrow}.

At the core of Arrow is a columnar memory format for flat and hierarchical data. This
format enables (1) fast analytical data processing and vectorized execution, and (2)
zero-deserialization data interchange. To achieve the former, Arrow organizes data
contiguously in 8-byte aligned buffers and uses separate bitmaps for nulls. For the
latter, Arrow specifies a standard in-memory representation and provides a C-like data
definition language (DDL) for data schema. Arrow uses separate metadata data structures
to impose a table-like structure on collections of buffers. An example of this for the
TPC-C \dbTable{ITEM} table is shown in \cref{fig:arrow-ddl}.

\begin{figure}[t!]
    \centering
    \includegraphics[width=\columnwidth]{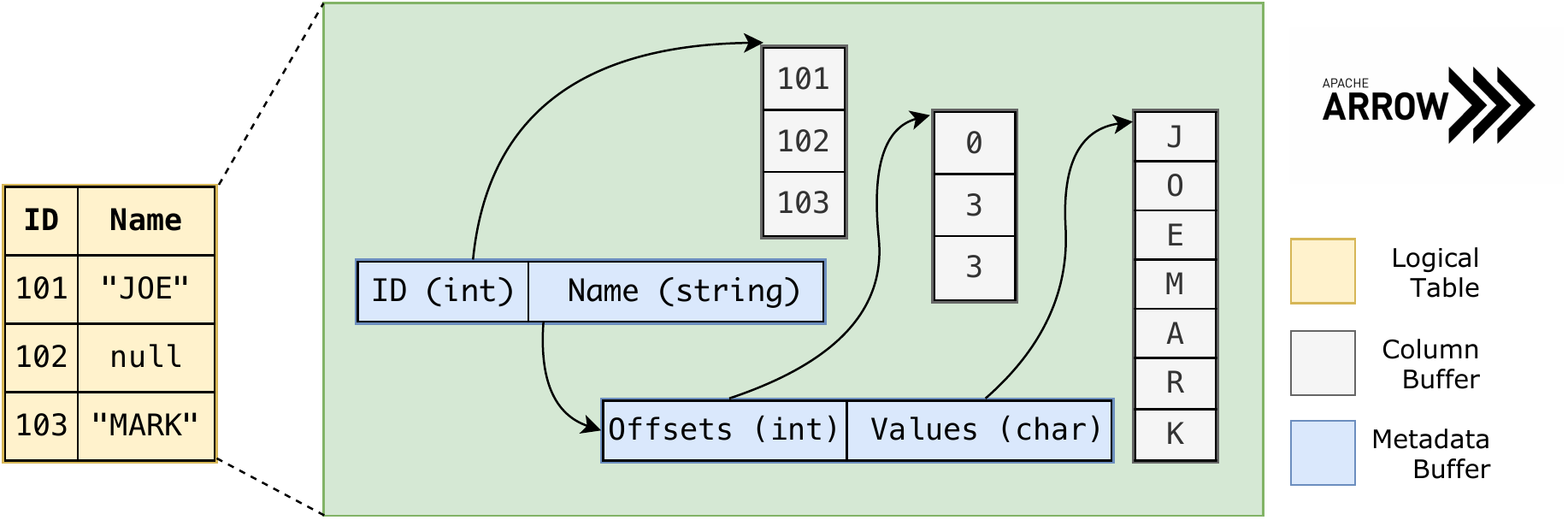}
    \caption{
        \textbf{Variable Length Values in Arrow} -- 
        Arrow represents variable length values as an offsets array
        into an array of bytes, which trades off efficient mutability\
        for read performance.
    }
    \label{fig:arrow-varlen}
\end{figure}

Although Arrow's design targets read-only analytical workloads, its alignment requirement
and null bitmaps also benefit write-heavy workloads on fixed-length values. Problems
emerge in Arrow's support for variable-length values (e.g., \texttt{VARCHAR}s). Arrow
stores them as an array of offsets indexing into a contiguous byte buffer. As shown in
\cref{fig:arrow-varlen}, the values' lengths are the difference between their starting
offset and the next value. This approach is not ideal for updates because of write
amplification. Suppose a program update the value ``\texttt{JOE}'' to ``\texttt{ANNA}''
in \cref{fig:arrow-varlen}, it must copy the entire \texttt{Values} buffer to a larger
one and update the \texttt{Offsets} array. The core of this issue is that a single
storage format cannot easily achieve simultaneously (1) data locality and value adjacency,
(2) constant-time random access, and (3) mutability~\cite{athanassoulis16}, which Arrow
trades off.

Some researchers have proposed hybrid storage schemes of row-store and column-store to 
get around this trade-off. Two notable examples are \peloton~\cite{arulraj16} and
\htwoo~\cite{h2o}. \peloton uses an abstraction layer above the storage engine that
transforms cold row-oriented data into a columnar format. In contrast, \htwoo uses an
abstraction layer at the physical operator level and generates code for the optimal
format on a per-query basis. Both solutions see an increase in software engineering
complexity, and limited speedup in the OLTP scenario (shown in \cref{sec:eval-oltp}).
We therefore argue that while it makes sense to optimize the data layout differently
based on access patterns, column-stores are good enough for both OLTP and OLAP use
cases. 

% We show in
% \cref{sec:transformation} how we design around this restriction by efficient converting
% between two similar designs that achieve (1) for read-heavy blocks and (3) for
% update-heavy blocks, respectively.

\section{System Overview}
\label{sec:sys}
We now present \sysname's architecture . We first discuss how the
DBMS's \textit{\dbTxnManager} is minimally intrusive to Arrow's layout. We then describe
how it organizes tables into blocks and its addressing scheme for tuples. Lastly, we
describe the garbage collection and recovery components. For simplicity, we assume that
data is fixed length; we discuss variable-length data in the next section.

%% ---------------------------------------------------------------
%% Transaction Engine
%% ---------------------------------------------------------------
\subsection{Transactions on a Column-Store}
\label{sec:sys-txns}
\begin{figure}[t!]
    \centering
    \includegraphics[width=\columnwidth]{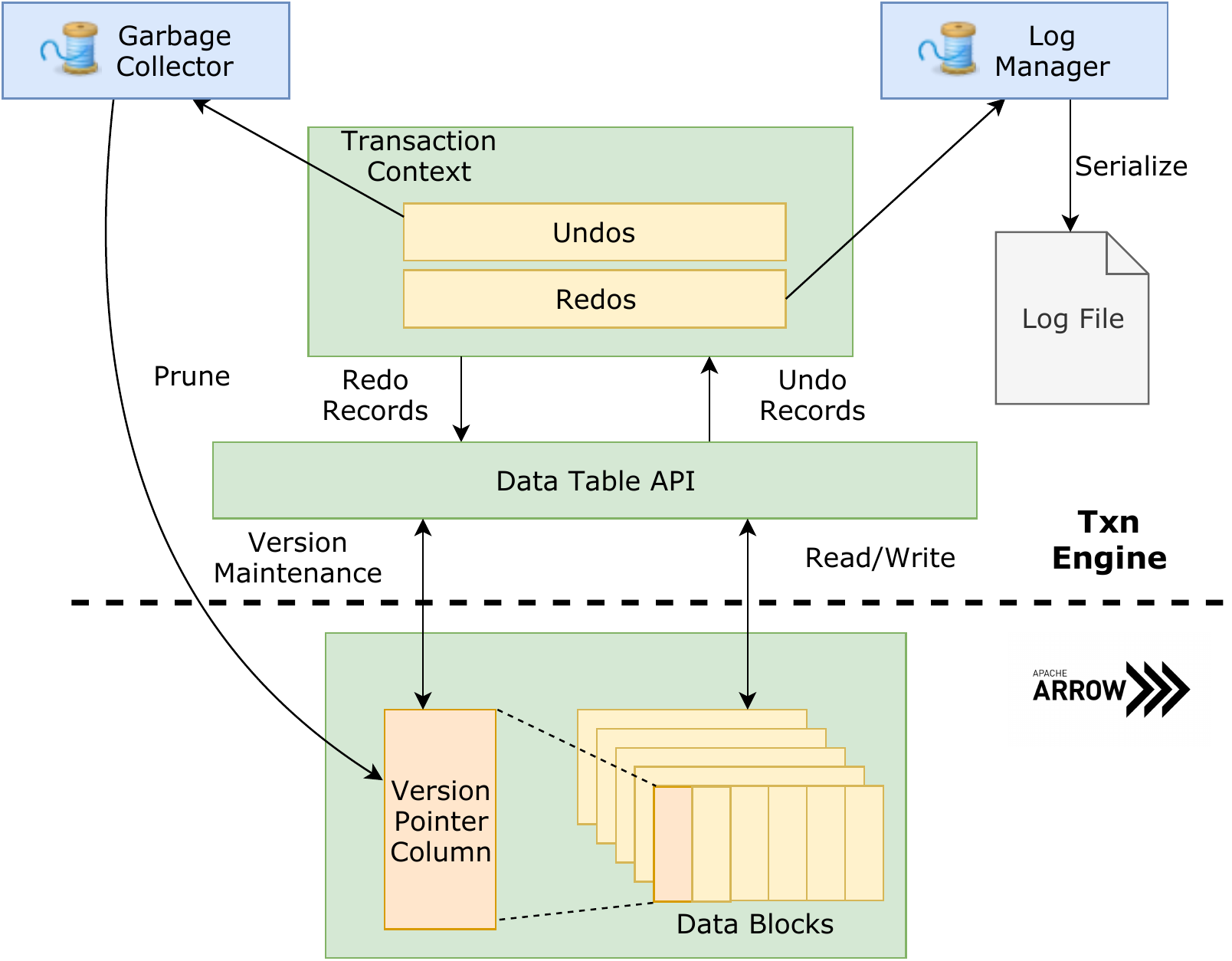}
    \caption{
        \textbf{System Architecture} -- 
        \sysname's transactional engine is minimally intrusive to the underlying storage to
        maintain compatibility with the Arrow storage format.
    }
    \label{fig:arch}
\end{figure}

An essential requirement for our system is that transactional and versioning metadata be
separate from the actual data; interleaving them complicates the mechanism for exposing
Arrow data to external tools. As shown in \cref{fig:arch}, our DBMS uses a
multi-versioned~\cite{bernstein83} delta-storage that stores the version chains as an
extra Arrow column that is invisible to external readers, where the system stores physical
pointers to the head of the version chain in the column (null if no version). The version
chain is a newest-to-oldest ordering of delta records, which are physical before-images
of the modified tuple attributes. This versioning approach enables the system to support Snapshot 
Isolation guarantees for concurrent transactions. 
The Data Table API serves as an abstraction layer to transactions, and will materialize
the correct version of the tuple into the transaction. This early materialization is
required for tuples with active versions, but can be elided for cold blocks, as we will
discuss in \cref{sec:transformation}. Version deltas are stored within the transaction
contexts, external to Arrow storage. The DBMS assigns each transaction an undo buffer as
an append-only row-store for deltas. To install an update, the transaction first reserves
space for a delta record at the end of its buffer, copies the current image of the modified 
attributes into the record, appends the record onto the version chain, and finally updates
the attribute in-place. The DBMS handles deletes and inserts analogously, but it updates a
tuple's allocation bitmap instead of its contents. This information is later passed to
the garbage collector and logging component of our system, as we discuss in
\cref{sec:sys-gc} and \cref{sec:sys-wal}.

A transaction's undo buffer needs to grow in size dynamically to support arbitrarily large
write sets. The DBMS cannot move delta records, however, as the version chain points
physically into the undo buffer. This rules out the use of a na\"{\i}ve resizing algorithm
that doubles the size of the buffer and copies the content. Instead, the system implements
undo buffers as a linked list of fixed-sized segments (currently 4096 bytes) and
incrementally adds new segments as needed.
% A 
% centralized object pool distributes buffer segments to transactions and recycles them when it is 
% safe. 

The \dbTxnManager assigns each transaction a timestamp pair (\tsStart, \tsCommit) that it
generates from the same counter. When a transaction starts, \tsCommit is the same as
\tsStart but with its sign bit flipped to denote that the transaction is uncommitted. Each
update on the version chain stores the transaction's \tsCommit timestamp. Readers
reconstruct their respective versions by copying the latest version, and then traversing
the version chain and applying before-images until it sees a timestamp less than its
\tsStart. Because the system uses unsigned comparison for timestamps, uncommitted versions
are never visible. The system disallows write-write conflicts to avoid cascading
rollbacks.

When a transaction commits, the DBMS uses a small critical section to obtain a commit
timestamp, update delta records' commit timestamps, and add them to the \dbLogManager's
queue. For aborts, the system uses the transaction's undo records to roll back the
in-place updates. It cannot unlink records from the version chain, however, due to
potential race conditions. If an active transaction copies a new version before the
aborting transaction that modified it performs the rollback, then the reader can traverse
the version chain with the undo record already unlinked and convince itself that the
aborted version is indeed visible. A simple check that the version pointer does not change 
while the reader makes a copy is insufficient in this scenario as the DBMS can encounter
an ``A-B-A'' problem between the two checks. To avoid this issue, the DBMS instead
restores the correct version before ``committing'' the undo record by flipping the sign
bit on the version's timestamp. This record is redundant for any readers that obtained the
correct copy and fixes the copy of readers with the aborted version.

Through this design, the \dbTxnManager reasons only about delta records and the version
column, and not the underlying physical storage. Maintaining the Arrow abstraction comes
at the cost of data locality and forces readers to materialize early, which degrades
range-scan performance. Fortunately, for many workloads, only a small fraction of the
database is expected to be versioned at any point in time. As a result, the DBMS can
ignore checking the version column for every tuple and scan large portions of the
database in-place~\cite{neumann15}. Blocks are natural units for tracking this
information, and the DBMS uses block-level locks to coordinate access to blocks that are
not versioned and not frequently updated (i.e., cold). We discuss this further in
\cref{sec:transformation}.

% We do not keep versioned indexes but instead model udpates as inserts and deletes into
% the index, which stores tuple slots as values. We clean up any stale entries in the
% index at the end of each transaction. \todo{fill in more detail as we implement indexes}
%% ---------------------------------------------------------------
%% Blocks
%% ---------------------------------------------------------------
\subsection{Blocks and Physiological Identifiers}
\label{sec:sys-blocks}
Separating tuples and transactional metadata introduces another challenge: the system
requires globally unique tuple identifiers to associate the two pieces that are not
co-located. Physical identifiers (e.g., pointers) are ideal for performance, but work
poorly with column-stores because a tuple does not physically exist at a single location.
Logical identifiers, on the other hand, must be translated into a memory location through
a lookup (e.g., hash table). This translation step is a severe bottleneck for OLTP
workloads because it potentially doubles the number of memory accesses per tuple.

To solve this, our DBMS organizes storage in 1~MB blocks, and uses a physiological scheme
to identify tuples. The DBMS arranges data in each block similar to PAX~\cite{pax}, where
all attributes of a tuple are within the same block. Every block has a layout object that
consists of (1) the number of slots within a block, (2) a list of attributes sizes, and
(3) the location offset for each column from the head of the block. Each column and its
bitmap are aligned at 8-byte boundaries. The system calculates layout once for a table
when the application creates it and uses it to handle every block in the table.

% In PAX, having an
% entire tuple located within a single block reduces the number of disk accesses on a write
% operations. Such benefits are less significant in an in-memory system; instead our system
% uses blocks as a useful logical grouping for tuples that are inserted close to each other
% chronologically that the system can use as a unit for transformation. Having all
% attributes of a tuple located within a single block also ensures that tuple will never be
% partially available in Arrow format.

Every tuple in the system is identified by a \dbTupleSlot, which is a combination of (1)
physical memory address of the block containing the tuple, and (2) its logical offset in
the block. Combining these with the pre-calculated block layout, the DBMS computes the
physical pointer to each attribute in constant time. To pack both values into a single
64-bit value, we use the C++11 keyword \texttt{alignas} to instruct the system to align
all blocks at 1 MB boundaries within the address space of the process. A pointer to a
block will then always have its lower 20 bits be zero, which the system uses to store the
offset. There are enough bits because there can never be more tuples than there are bytes
in a block.

% It is possible to further optimize
% this process by using a custom allocator specialized for handing out 1 MB chunks. The
% system can also store extra information in the address in a similar fashion; because it
% reserve space within each block for headers and version information, and because x86
% machines currently do not utilize the full 64 bits in a pointer, there are many bits to
% spare in a tuple slot. This allows the system to pack much information into the tuple
% identifier while keeping it in register for good performance when passing one around.
\begin{figure}[t!]
    \centering
    \includegraphics[width=0.89\columnwidth]{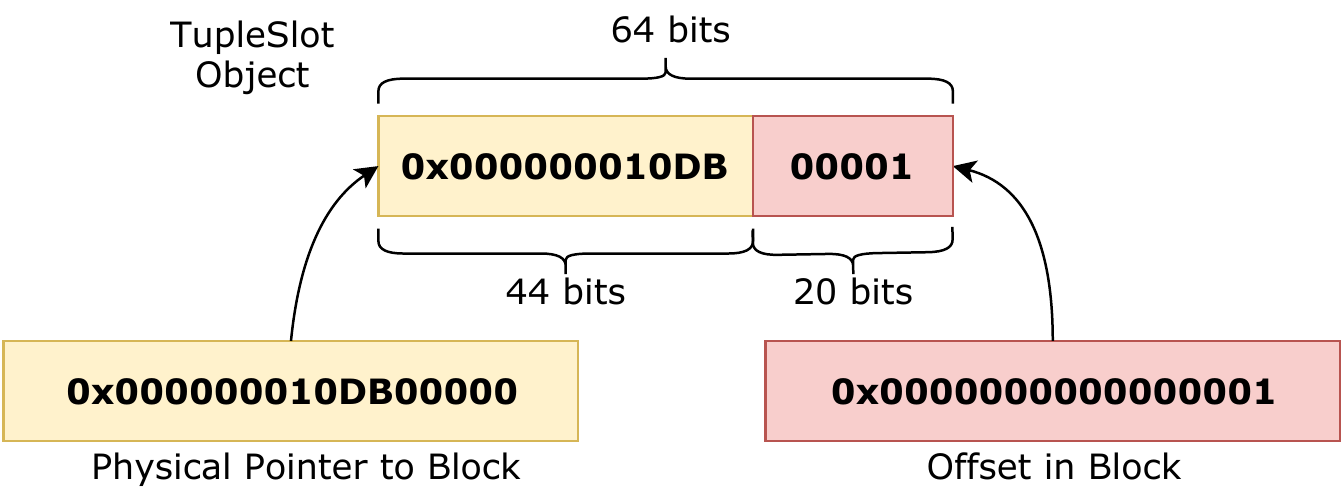}
    \caption{
        \textbf{TupleSlot} -- 
        By aligning blocks at 1 MB boundaries, the DBMS packs the
        pointer to the block and the offset in a 64-bit word.
    }
    \label{fig:tupleslot}
\end{figure}

%% ---------------------------------------------------------------
%% Garbage Collection
%% ---------------------------------------------------------------
\subsection{Garbage Collection}
\label{sec:sys-gc}
In our system, the garbage collector (GC)~\cite{larson11,tu13, yu14, lee16} is responsible for 
pruning version chains and freeing any associated memory. The DBMS handles the recycling of deleted 
slots during the transformation to Arrow (\cref{sec:transformation-algo}). Because the DBMS 
stores versioning information within a transaction's buffers, the GC only examines transaction 
objects.

At the start of each run, the GC first checks the \dbTxnManager's transactions table for the oldest 
active transaction's \tsStart timestamp; changes from transactions committed before this 
timestamp are no longer visible and are safe for removal. The GC inspects all such transactions to 
compute the set of \dbTupleSlot{s} that have invisible records in their version chains, and 
then truncates them exactly once. This step avoids the quadratic operation of finding and unlinking 
each record. 
Deallocating objects is unsafe at this point, however, as concurrent transactions may be reading 
the unlinked records. To address this, GC obtains a timestamp from the \dbTxnManager that represents 
the time of unlink. Any transaction starting after this time cannot possibly access the unlinked 
record; 
the records are safe for deallocation when the oldest running transaction in the system has a larger 
\tsStart timestamp than the unlink time. Our approach is similar to an epoch-protection 
mechanism~\cite{chandramouli18}, and is generalizable to ensure thread-safety for other aspects of 
the DBMS as well.

%The GC process traverses every update exactly once and requires no additional full table 
%scans. %  and easily keeps up with high throughput.
%The combination of these two facets gives us leeway to incorporate
%additional checks on each garbage collector run without impacting performance.
%We take advantage of this to compute access statistics about blocks, as we will describe in 
%\cref{sec:transformation-coldblocks}.

%% ---------------------------------------------------------------
%% Logging and Recovery
%% ---------------------------------------------------------------
\subsection{Logging and Recovery}
\label{sec:sys-wal}
Our system achieves durability through write-ahead logging and
checkpoints~\cite{mohan92, dewitt84}. Logging in the DBMS is analogous to the
GC process described above. Each transaction maintains a redo buffer for physical after-images. 
Each transaction writes changes to its redo buffer in the order that they occur.
At commit time, the transaction appends a commit record to its redo buffer and adds itself to the 
DBMS's
flush queue. The \dbLogManager asynchronously serializes the changes from these buffers into an 
on-disk format before flushing to persistent storage. The system relies on an implicit ordering of 
the records according to their respective transaction's \tsCommit timestamp instead of log sequence 
numbers.

% There are some caveats worth pointing out here. One is that the entries in the redo buffer
% are not marshalled bytes, but direct memory images of the updates, with paddings and
% swizzled pointers. The logging thread therefore has to serialize the records into their
% on-disk formats before flushing. This process happens off the critical path, however,
% and we have not found it to be a significant overhead in our evaluations.

Similar to undo buffers, these redo buffers consist of buffer segments drawn from a 
global object pool. As an optimization, the system flushes out redo records incrementally before
the transaction commits. In the case of an abort or crash, the transaction's commit
record is not written, and the recovery process ignores it. In our implementation, we
limit the redo buffer to a single buffer segment and observe moderate speedup due to
better cache performance from more reuse.

The rest of the system considers a transaction as committed as soon as its commit record
is added to the flush queue. All future operations on the transaction's write-set are
speculative until its log records are on disk. The system assigns a callback to each
committed transaction for the \dbLogManager to notify when the transaction is persistent.
The DBMS refrains from sending a transaction's result to the client until the
\dbLogManager invokes its callback. With this scheme, a transaction's modifications that
speculatively accessed or updated the write-set of another transaction are not published
until the \dbLogManager processes their commit record. We implement callbacks by embedding
a function pointer in the commit record; when the \dbLogManager writes the commit record,
it adds that pointer to a list of callbacks to invoke after the next \texttt{fsync}. The
DBMS requires read-only transactions also to obtain a commit record to guard against the
anomaly shown above. The \dbLogManager can skip writing this record to disk after
processing the callback.

\section{Block Transformation}
\label{sec:transformation}
As discussed in \cref{sec:bg-arrow}, the primary obstacle to running transactions on Arrow
is write amplification. Our system uses a relaxed Arrow format to achieve good write
performance and then uses a lightweight transformation step to put a block into the full
Arrow format once it is cold. In this section, we describe this modified format, introduce
a mechanism to detect cold blocks and present our algorithm for transforming them to the
full Arrow format. 

\begin{figure}[t!]
    \centering
    \includegraphics[width=\columnwidth]{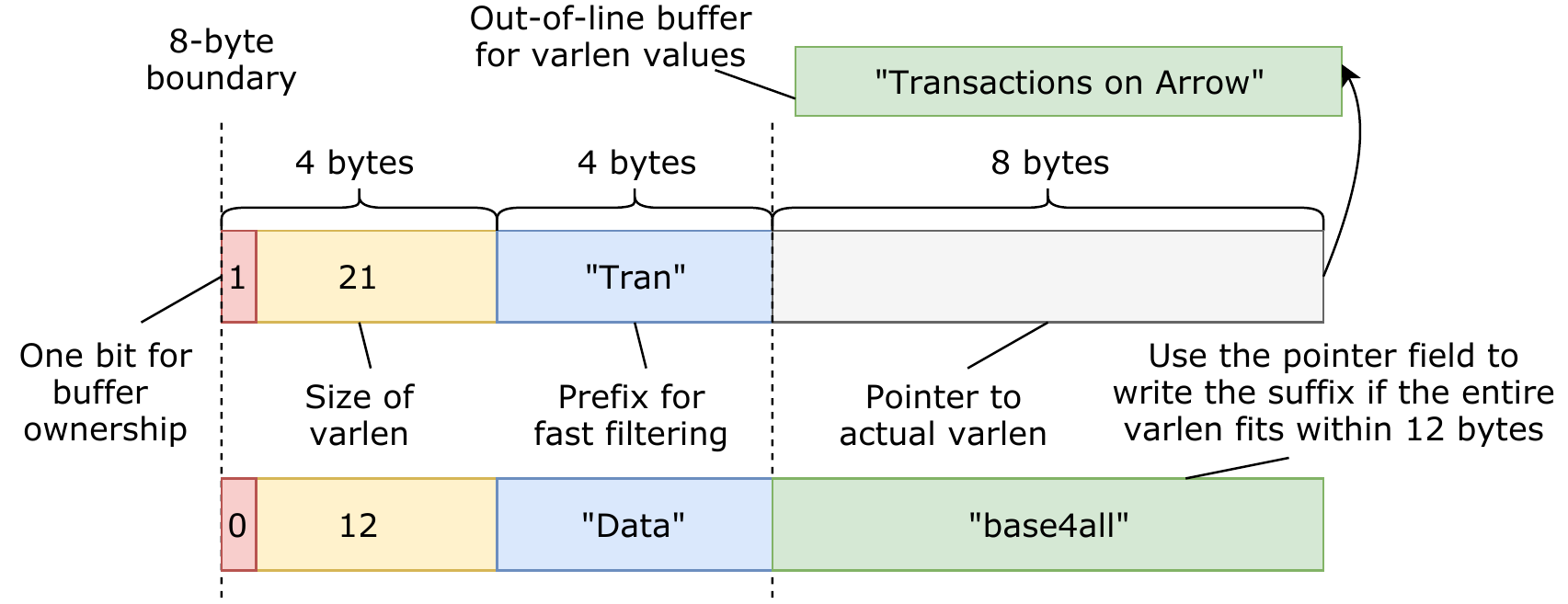}
    \caption{
        \textbf{Variable-Length Value Storage} -- 
        The system stores variable-length values as a 16-byte column in a block.
    }
    \label{fig:varlen-entry}
\end{figure}

%% ---------------------------------------------------------------
%% Relaxed Columnar Format
%% ---------------------------------------------------------------
\subsection{Relaxed Columnar Format}
\label{sec:transformation-format}
Typical OLTP workloads modify only a small portion of a database at any given time, while
the other parts of the database are mostly accessed by read-only queries~\cite{lang}.
Therefore, for the hot portion, we can trade off read speed for write performance at only
a small impact on overall read performance of the DBMS. To achieve this, we modify the
Arrow format for update performance in the hot portion. We detail these changes in this
subsection.
% This forces all readers, both
% internal and external, to access data via the slower path that materializes versions. We
% contend that the cost is acceptable as this materialization happens only for a small
% portion of the database.

\begin{figure}[t!]
    \centering
    \includegraphics[width=\columnwidth]{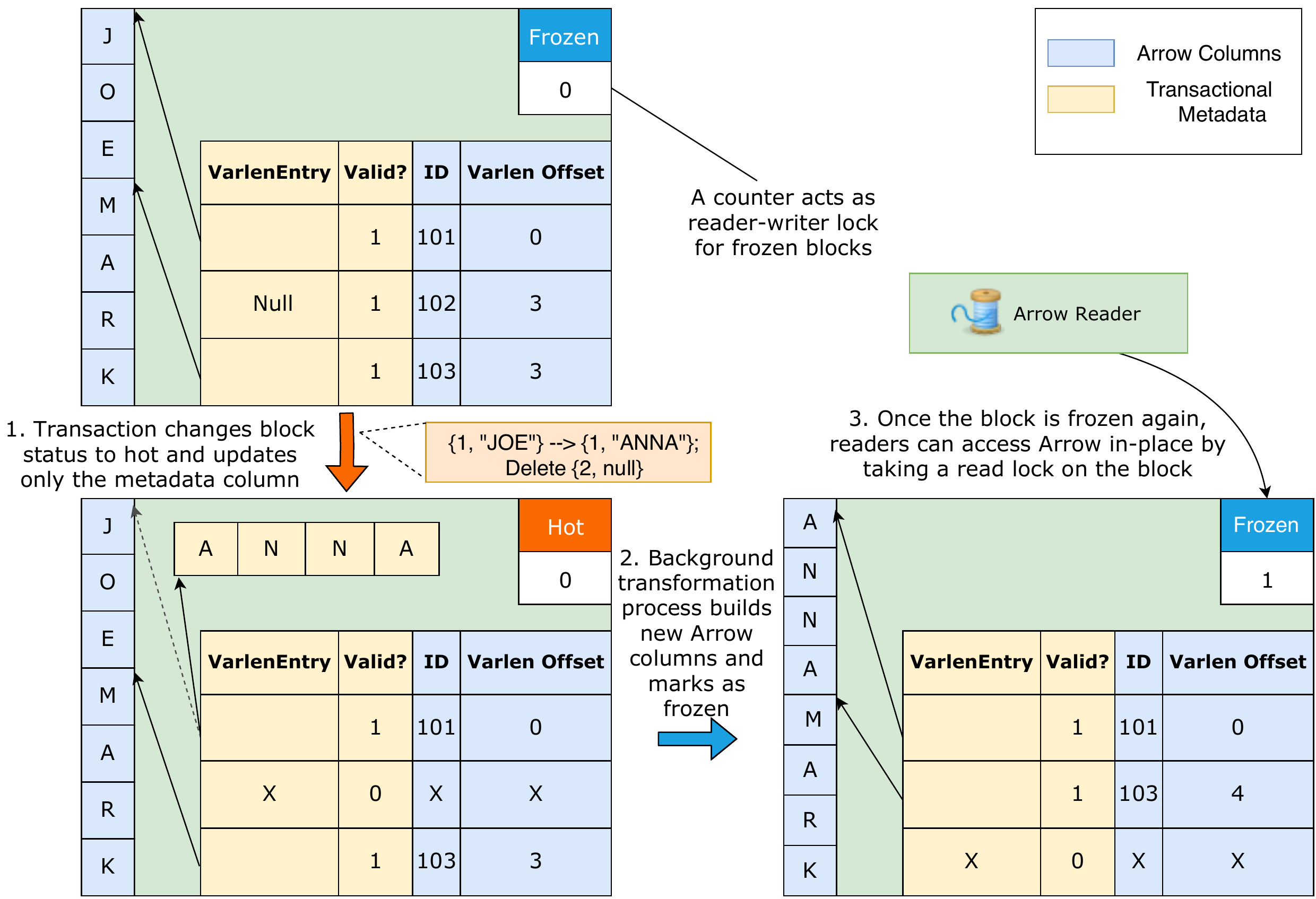}
    \caption{
        \textbf{Relaxed Columnar Format} -- 
        The system briefly allows non-contiguous memory to support
        efficient updates.
    }
    \label{fig:format}
\end{figure}

There are two sources of write amplification in Arrow: (1) it disallows gaps in a column,
and (2) it stores variable-length values consecutively in a single buffer. Our relaxed
format adds a validity bitmap in the block header and additional metadata for each
variable-length value in the system to overcome them. As shown in \cref{fig:varlen-entry},
within a \dbVarlenEntry field, the system maintains 4 bytes for size and 8 bytes for a
pointer to the underlying value. Each \dbVarlenEntry is padded to 16 bytes for alignment
reasons, and the additional 4 bytes stores a prefix of the value. If a value is shorter
than 12 bytes, the system stores it entirely within the object, writing into the pointer.
Transactions only access the \dbVarlenEntry instead of Arrow storage directly. Relaxing
adherence to Arrow's format allows the system to only write updates to \dbVarlenEntry,
turning a variable-length update into a constant-time fixed-length one, as shown in
\cref{fig:format}.

Any readers accessing Arrow storage will be oblivious to the update in \dbVarlenEntry. The
system adds a \textit{status flag} and \textit{counter} in block headers to coordinate
access. For a cold block, the DBMS sets its status flag to \textbf{\statusFrozen}, and
readers add one to the counter when starting a scan and subtract one when finished. When a
transaction updates a cold block, it first sets that block's status flag to
\textbf{\statusHot}, forcing any future readers to materialize instead of reading
in-place. It then spins on the counter and waits for lingering readers to leave the block
before proceeding with the update. There is no transformation process required for a
transaction to modify a cold block because our relaxed format is a super-set of the
original Arrow format. Once a block is hot, it remains so until a background process
transforms the block back to full Arrow compliance. We will discuss this process next.

%% ---------------------------------------------------------------
%% Identifying a Cold Block
%% ---------------------------------------------------------------
\subsection{Identifying Cold Blocks}
\label{sec:transformation-coldblocks}
The DBMS maintains access statistics about each block to determine if it is cooling down.
Collecting them as transactions operate on the database adds overhead to the critical 
path~\cite{hyper12,debrabant13}, which is unacceptable for OLTP workloads. Our system
trades the quality of such statistics for better scalability and performance, and then
accounts for potential mistakes from this in our transformation algorithm.

A simple heuristic is to mark blocks that have not been modified for some threshold time
as cold for each table. Instead of measuring this on the transaction's critical path, our
system takes advantage of the GC's scan through undo records (\cref{sec:sys-gc}). From
each undo record, the system obtains the modification type (i.e., delete, insert, update)
and the corresponding \dbTupleSlot. Time measurement, however, is difficult because the
system cannot measure how much time has elapsed between the modification and invocation of
the GC. The DBMS instead approximates this by using the time of each GC invocation as the
time for the modifications processed in said GC run. If transactions have a lifetime
shorter than the frequency of GC ($\sim$10~ms), this approximated time is never earlier
than the actual modification and is late by at most one GC period. This ``GC epoch'' is a
good enough substitute for an exact time for short-lived OLTP
transactions~\cite{stonebraker2007}. Once the system identifies a cold block, it adds the
block to a queue for background processing. The user can modify the threshold time value
based on how aggressively they want the system to transform blocks. The optimal value is
workload-dependent. A threshold that is too low reduces transactional performance because
of wasted resources from frequent transformations. But setting it too high reduces the
efficiency of readers. We leave the study of more sophisticated policies for future work.

Under this scheme, one thread may identify a block as cold by mistake when another thread
is updating it due to delays in access observation. The DBMS reduces the impact of this by
ensuring that the transformation algorithm is fast and lightweight. There are two failure
cases: (1) a user transaction aborts due to conflicts with the transformation process or
(2) the user transaction stalls. There is no way to safely eliminate both cases. Our
solution is a two-phase algorithm. The first phase is transactional and operates on a
microsecond scale, minimizing the possibility of aborts. The second phase eventually takes
a block-level lock for a short critical section, but yields to user transactions whenever
possible.

%% ---------------------------------------------------------------
%% Transformation Algorithm
%% ---------------------------------------------------------------
\subsection{Transformation Algorithm}
\label{sec:transformation-algo}
\begin{figure}[t!]
    \centering
    \includegraphics[width=\columnwidth]{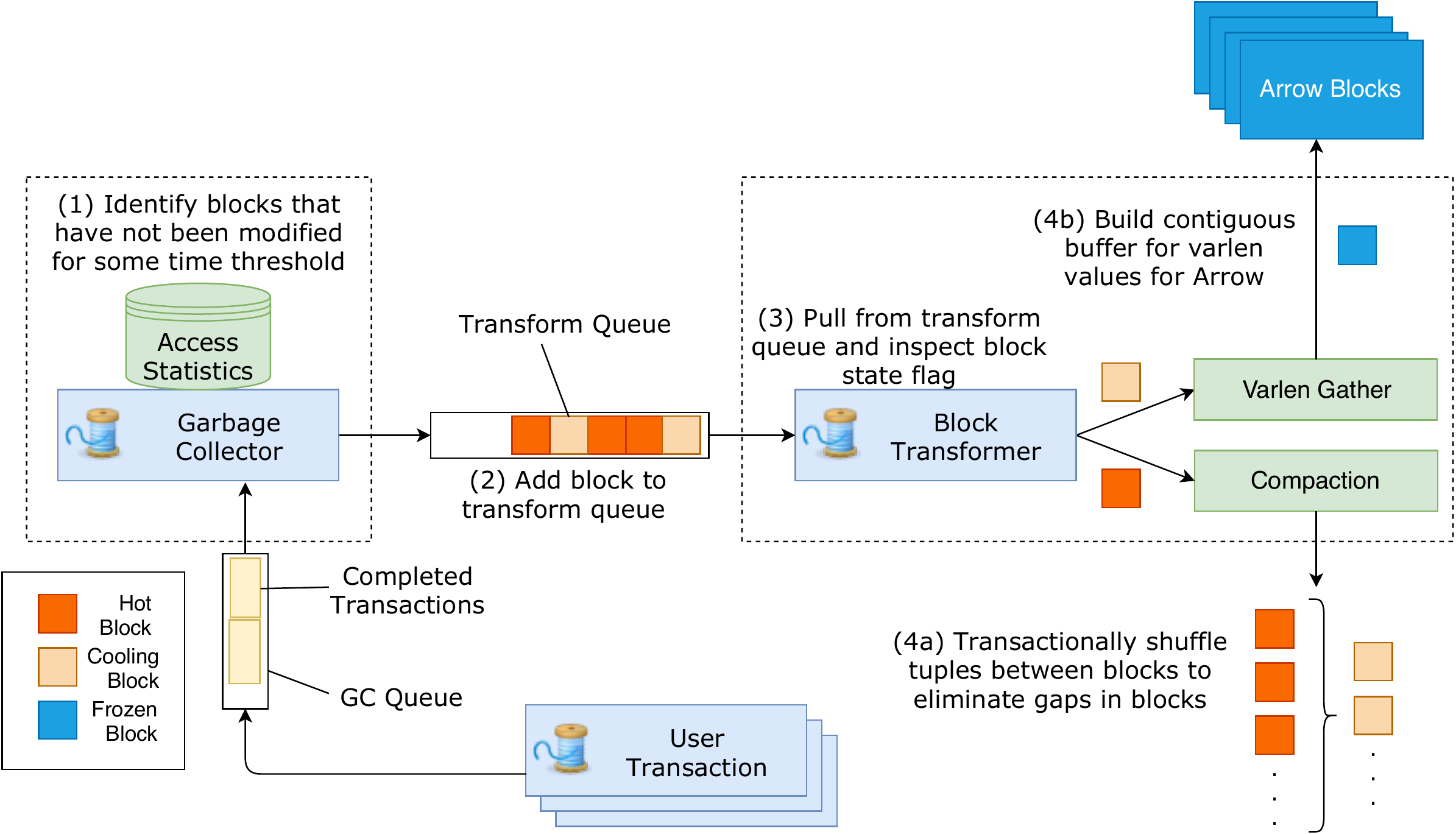}
    \caption{
        \textbf{Transformation to Arrow} -- 
        \sysname implements a pipeline for lightweight in-memory transformation to Arrow.
    }
    \label{fig:algo}
\end{figure}

Once the system identifies cooling blocks, it performs a transformation pass to prepare
the block for Arrow readers. As mentioned in \cref{sec:transformation-format}, the DBMS
first needs to compact each block to eliminate any gaps, and then copy variable-length
values into a new contiguous buffer. There are three approaches to ensure safety in the
face of concurrent user transactions: (1) copying the block, (2) performing operations
transactionally, or (3) taking a block-level lock. None of these is ideal. The first
approach is expensive, especially when most of the block data is not changed. The second
adds additional overhead and results in user transaction aborts. The third stalls user
transactions and limits concurrency in the typical case even without transformation. As
shown in \cref{fig:algo}, our system uses a hybrid two-phase approach that combines
transactional tuple movement and raw operations under exclusive access. We now discuss
this in detail.
\\ \vspace{-0.1in}

%% -----------------------
%% Compaction
%% -----------------------
\textbf{Phase \#1: Compaction:}
The access observer identifies a \textit{compaction group} as a collection of blocks with
the same layout to transform. Within a group, the system uses tuples from less-than-full
blocks to fill gaps in others and recycle blocks when they become empty. The DBMS uses one
transaction per group in this phase to perform all operations. 

The DBMS scans the allocation bitmap of every block to identify empty slots that it 
needs to fill. At the end of this phase, tuples in the compaction group should be
``logically contiguous'', i.e., a compaction group consisting of $t$ tuples with $b$
blocks with each block having $s$ slots should now have
$\left\lfloor{\frac{t}{s}}\right\rfloor$ many blocks completely filled, one block filled 
from beginning to the $(t\!\!\mod\!s)$-th slot, and all remaining blocks empty. To achieve
this, the system transactionally shuffles tuples between blocks (delete followed by an
insert). This is potentially expensive if the transaction needs to update indexes. The
algorithm, therefore, must minimize the number of such delete-insert pairs. We do this
in two steps:

\squishlist
    \item
    Select a block set $F$ to be the $\left\lfloor{\frac{t}{s}}\right\rfloor$
    blocks that are filled in the final state. Also select a block $p$ to be
    partially filled and hold $t\!\!\mod\!s$ tuples. The rest, $E$, are left
    empty.

    \item
    Fill all gaps within $F \cup \{p\}$ using tuples from $E \cup \{p\}$, and
    reorder tuples within $p$ to make them contiguous.
\squishend

Let $Gap_f$ be the set of unfilled slots in a block $f$, $Gap'_f$ be the set of
unfilled slots in the first $t\!\!\mod\!s$ slots in a block $f$, $Filled_f$ be the set of
filled slots in $f$, and $Filled'_f$ be the set of filled slots not in the first
$t\!\!\mod\!s$ slots in $f$. Then, for any valid selection of $F$, $p$, and $E$,
$$|Gap'_p| + \Sigma_{f \in F}|Gap_f| = |Filled'_p| + \Sigma_{e \in E}|Filled_e|$$
because there are only $t$ tuples in total. Therefore, given $F$, $p$, and $E$, an optimal
movement is any one-to-one movement between $Filled'_p\ \cup\ \bigcup_{e \in E}Filled_e$
and $Gap'_p\ \cup\ \bigcup_{f \in F}Gap_f$. The problem is now reduced to finding such
$F$, $p$ and $E$.
\squishlist
  \item Scan each block's allocation bitmap for empty slots.
  \item Sort the blocks by \# of empty slots in ascending order.
  \item Pick out the first $\left\lfloor{\frac{t}{s}}\right\rfloor$ blocks to be $F$.
  \item Pick an arbitrary block as $p$ and the rest as $E$.
\squishend

This choice bounds our algorithm to within $(t\!\!\mod\!s)$ of the optimal number of
movements, which we use as an approximate solution.  Every gap in $F$ needs to be
filled with one movement, and our selection of $F$ results in fewer movements than
any other choice. In the worst case, the chosen $p$ is empty in the first
$(t\!\!\mod\!s)$ slots, and the optimal one is filled, resulting in at most
$(t\!\!\mod\!s)$ movements of difference. The algorithm needs to additionally find the
best value of $p$ by trying every block for the optimal solution. In practice,
%the system
%can choose the block with fewest unfilled slots not in $F$ to be $p$, and the worst case
%scenario is unlikely to happen. 
as described in \cref{sec:eval}, we observe only marginal reduction in movements,
which does not always justify the extra step. 
\\ \vspace{-0.1in}

%% -----------------------
%% Gathering
%% -----------------------
\textbf{Phase \#2: Gathering:}
The system now moves variable-length values into contiguous buffers as Arrow requires. 
%As we have discussed previously, neither transactional
% updates nor block locks are efficient enough by themselves. Therefore,
To do so safely, we present a novel scheme of multi-stage locking that relies on the GC
to guard against races without requiring other operations to obtain the lock.

% In the na\"{\i}ve implementation, it suffices to continue
% updating the table in the same transaction used by the compaction phase. The system
% allocates a buffer, copies scattered variable length values into it, and updates the
% tuple's \dbVarlenEntry columns to point to the new buffer. Eventually, either
% the transformation transaction is aborted due to conflicts, or it succeeds in updating
% every tuple in the blocks being transformed. Because our system is no-wait, any other
% transaction that attempt to update values in those blocks will be aborted, and the
% correctness of the operation is guaranteed.

% Unfortunately, as we will demonstrate in \cref{sec:eval}, this na\"{\i}ve approach has
% suboptimal performance. This is because transactional updates do additional work to ensure
% isolation; transactional updates have irregular memory access patterns compared to
% sequential in-place updates, and uses the expensive compare-and-swap instruction for every
% tuple. Because we already assume the blocks being processed here are no longer
% transactionally updated, transactions incur much overhead to guard against contention
% that rarely happens. Therefore, to speed up the gathering phase further, it is necessary
% to elide transactional protection and use locking for the duration of the operation. 
% \todo{need experimental evidence} That said, introducing a shared lock at the block
% level may slow down the common case scenario as well. To achieve the best of both
% worlds, 

We extend the block status flag with two additional values: \statusCooling and
\statusFreezing. The former indicates that the transformation thread intends to lock,
while the latter serves as an exclusive lock that blocks user transactions. User
transactions are allowed to preempt the \statusCooling status by compare-and-swapping the
flag back to \statusHot. When the transformation algorithm has finished compaction, it
sets the flag to \statusCooling and scans through the block again to check for any version
pointers, which indicate concurrent modification. If there are no versions, and another
thread has not changed the block's \statusCooling status, then the transformation
algorithm can change the block's status to \statusFreezing for the exclusive lock.
The \statusCooling flag acts as a sentinel value that detects any concurrent modifications
that the single-pass scan missed.

\begin{figure}[t!]
    \centering
    \includegraphics[width=0.75\columnwidth]{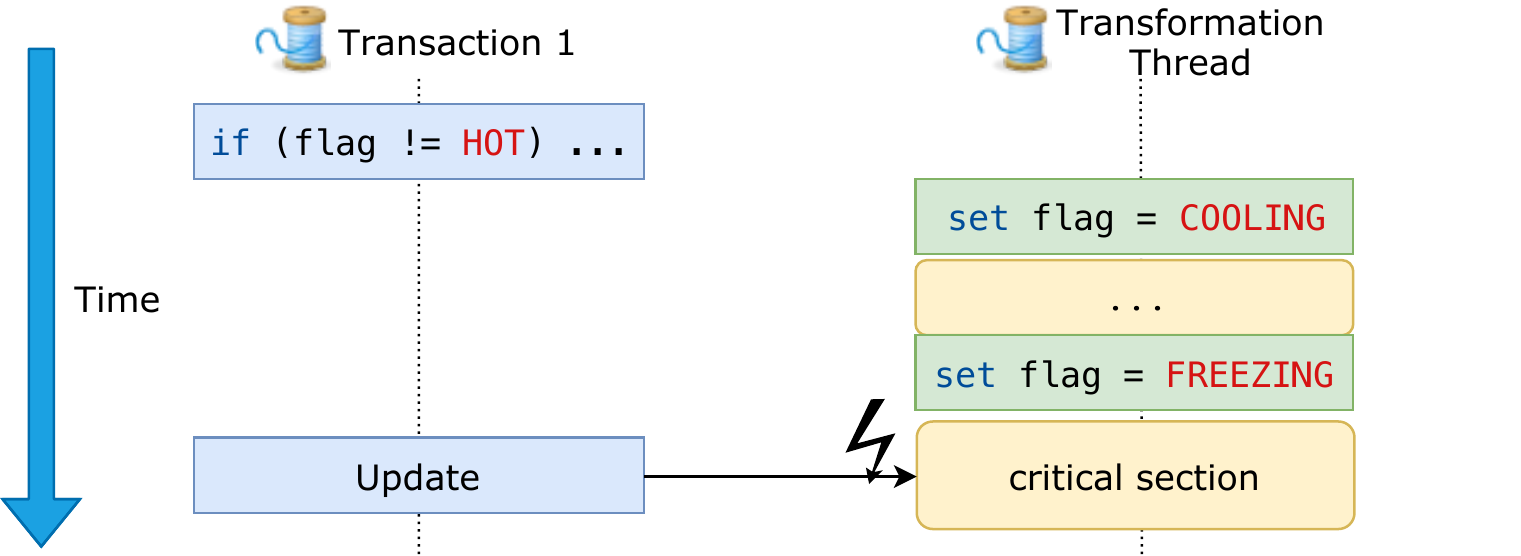}
    \caption{
        \textbf{Check-and-Miss on Block Status} -- 
        A na\"{\i}ve implementation results in a race during the gathering phase.
    }
    \label{fig:race}
\end{figure}

This scheme of access coordination introduces a race, as shown in \cref{fig:race}. A
thread could have finished checking the status flag and was scheduled out. Meanwhile, the
transformation algorithm runs and sets the block to \statusFreezing. When the thread wakes
up again, it proceeds to update, which is unsafe. The core issue here is that the block
status check and the update form a critical section but cannot be atomic without a latch.
Adding a latch for every operation is clearly undesirable. To address this, the system
relies on its visibility guarantees. Recall from \cref{sec:sys-gc} that GC does not prune
any versions that are still visible to running transactions. If the algorithm sets the
status flag to \statusCooling after shuffling, but before the compaction transaction
commits, the only transactions that could incur the race in \cref{fig:race} must overlap
with the compaction transaction. Therefore, as long as such transactions are alive, the
garbage collector cannot unlink records of the compaction transaction. The algorithm can
commit the compaction transaction and wait for the block to reappear in the processing
queue for transformation. The status flag of \statusCooling guards against any
transactions modifying the block after the compaction transaction committed. If the
transformation algorithm scans the version pointer column and finds no active version,
then any transaction that was active at the same time as the compaction transaction must
have ended, and it is safe to change the flag into \statusFreezing.

After the transformation algorithm obtains exclusive access to the block, it scans
each variable-length column and performs gathering in-place. In the same pass, it also
computes metadata information, such as null count, for Arrow's metadata. When the process
is complete, the system can safely mark the block as \statusFrozen and allow access from
in-place readers. Throughout the process, although transactional writes are not allowed,
reads can still proceed regardless of the block status. The gathering phase changes only
the physical location of values and not the logical content of the table. Because a write
to any aligned 8-byte address is atomic on a modern architecture~\cite{intel-manual},
reads can never be unsafe as the DBMS aligns all attributes within a block.

%% ---------------------------------------------------------------
%% Transformation Algorithm
%% ---------------------------------------------------------------
\subsection{Additional Considerations}
\label{sec:transformation-additional}
Given that we have presented our algorithm for transforming cold blocks into Arrow, we now
demonstrate its flexibility by discussing alternative formats for our transformation
algorithm with the example of dictionary compression. We also give a more detailed
description of memory management in the algorithm and scaling for larger workloads.
\\ \vspace{-0.1in}

%% -----------------------
%% Dictionary Encoding
%% -----------------------
\textbf{Alternative Formats:}
It is possible to change the implementation of the gathering phase to emit a different
format, although the algorithm performs best if the target format is close to our
transactional representation. To illustrate this, we implement an alternative columnar format with 
the same kind of dictionary compression~\cite{Holloway08} found in formats like 
Parquet~\cite{parquet} and ORC~\cite{apacheorc}.
%Dictionary compression is a compression technique used
%in read-optimized databases that replaces attribute values with an offset into a value
%table~\cite{Holloway08}.
Instead of building a contiguous variable-length buffer, the system creates a dictionary
and an array of dictionary codes. Much of the algorithm remains the same; the only
difference is that within the critical section of the gathering phase, the algorithm now
scans through the block twice. On the first scan, the algorithm builds a sorted set of
values for use as a dictionary. On the second scan, the algorithm replaces pointers within
\dbVarlenEntry{s} to point to the corresponding dictionary word and builds the array of
dictionary codes. Although the steps for transforming data into this format is mostly the 
same as Arrow, supporting dictionary compression is an
order of magnitude more expensive than a simple variable-length gather. We discuss the
effect of this procedure in \cref{sec:eval}.
% Some other possible targets include emitting Apache Parquet or simple data preparation tasks. 
\\ \vspace{-0.1in}

%% -----------------------
%% Memory Management
%% -----------------------
\textbf{Memory Management:}
Since the algorithm never blocks readers, the system cannot deallocate memory immediately
after the transformation process as its content can be visible to concurrent transactions.
In the compaction phase, because writes are transactional, the GC can handle memory
management. The only caveat here is that when moving tuples, the system makes a copy of
any variable-length value rather than merely copying the pointer. This value copy is
necessary because the GC does not reason about the transfer of ownership of
variable-length values between two versions and will deallocate them after seeing the
deleted tuple. We do not observe this to be a bottleneck. In the gathering phase, we
extend our GC to accept arbitrary actions associated with a timestamp in the form of a
callback, which it promises to invoke after the oldest alive transaction in the system is
started after the given timestamp. As discussed in \cref{sec:sys-gc}, this is similar to
an epoch protection framework~\cite{chandramouli18}. The system registers an action that
reclaims memory for this gathering phase with a timestamp that the compaction thread takes
after it completes all of its in-place modifications. This delayed reclamation ensures no
transaction reads freed memory.
\\ \vspace{-0.1in}

%% -----------------------
%% Scaling Transformation
%% -----------------------
\textbf{Scaling Transformation and GC:}
For high-throughput workloads (i.e., millions of transactions per second), a single GC or 
transformation thread will not be able to keep up. In this scenario, there is a natural partitioning 
of these tasks to enable parallelization. For GC, multiple threads can partition work based on 
transactions: when a transaction finishes, the DBMS assigns its clean-up operations to a random GC 
thread or according to some other load-balancing scheme. Although the pruning of version chain 
itself is thread-safe, multiple GC threads pruning the same version chain can do so at a different 
pace, and deallocate parts of the chain in the other's path. This concurrency is also wasteful as a 
version chain only needs to be pruned once every GC invocation. Therefore, in our system, GC threads 
mark the head of a version chain when pruning as a signal for others to back off. To parallelize 
transformation, the DBMS can partition threads on a compaction group level. No changes to the 
transformation process are required, as compaction groups are isolated units of work that never 
interfere with each other. We use these two techniques in \cref{sec:eval-oltp} on high worker thread 
counts. 

% ==================================================================
%% EXTERNAL ACCESS
%% ==================================================================
\section{External Access}
\label{sec:export}
Now that we have described how the DBMS converts data blocks into the Arrow format, we
discuss how to expose access to external applications. We argue that native Arrow storage
can benefit data pipeline builders, regardless of whether they take a
``data ships to compute'' approach or the opposite. In this section, we present three
strategies for enabling applications to access the DBMS's native Arrow storage to speed up
analytical pipelines. We discuss these alternatives in the order of the engineering effort
required to change an existing system (from easiest to hardest).
\\ \vspace{-0.1in}

%% -----------------------
%% Wire Protocol
%% -----------------------
\textbf{Improved Wire Protocol:}
There are still good reasons for applications to interact with the DBMS exclusively
through a SQL interface (e.g., developer familiarity, existing ecosystems). As
\cite{raasveldt17} pointed out, adopting columnar batches instead of rows in the wire
format can increase performance substantially. Arrow data organized by block is naturally
amenable to such wire protocols. However, replacing the wire protocol with Arrow does not 
achieve the full potential of the speed-up from our storage scheme. This is because the
DBMS still serializes data into its wire format, and the client must parse the data. These
two steps are not necessary, as the client may want the Arrow format to work with anyway.
The DBMS should be able to send stored data directly onto the wire and land them in the
client program's workspace, without writing to or reading from a wire format. For this purpose, 
Arrow provides a native RPC framework based on gRPC called
Flight~\cite{apachearrow-github} that avoids serialization when transmitting
data, through low-level extensions to gRPC's internal memory management. Flight enables
our DBMS to send a large amount of cold data to the client in a zero-copy fashion. When
most data is cold, Flight transmits data significantly faster than real-world DBMS
protocols. To handle hot data, the system needs to start a transaction and materialize a
snapshot of the block before invoking Flight. Although this is expensive, we observe that
Flight still performs no worse than state-of-the-art protocols~\cite{raasveldt17}.
\\ \vspace{-0.1in}

%% -----------------------
%% RDMA
%% -----------------------
\textbf{Shipping Data with RDMA:}
To achieve further speed-up, one can consider Remote Direct Memory Access (RDMA)
technologies. RDMA bypasses the OS's network stack and permits high-throughput,
low-latency transfer of data. Either the client or the DBMS can RDMA into the other's
memory, and we sketch both scenarios. 

The DBMS server can write data to the client's memory through RDMA (i.e., client-side
RDMA). Under this scheme, the server retains control over access to its data, and no
modification to the concurrency control scheme is required. Aside from increased data
export speed, another benefit of using a client-side approach is that the client's CPU
is idle during RDMA operations. Thus, the client can start working on partially available
data, effectively pipelining data processing. To achieve this, the DBMS can send messages
for partial availability of data periodically to communicate whether it has already
written some given chunk of data. This approach reduces the network traffic close to its
theoretical lower-bound but still requires additional processing power on the server to
handle and service the request.

For workloads that require no computation on the server-side, allowing clients to read the
DBMS's memory (i.e., server-side RDMA) bypasses the DBMS CPU when satisfying bulk export
requests. This approach is beneficial to an OLTP DBMS because the system no longer needs
to divide its CPU resources between serving transactional workloads and bulk-export jobs.
Achieving server-side RDMA, however, requires significant changes to the DBMS. Firstly,
the DBMS loses control over access to its data as the client bypasses its CPU to get data
out, which makes it difficult to lock the Arrow block and guard against updates into them.
If the system waits for a separate client completion message, the round-trip time
introduces latency to any updating transactions. To avoid this, the DBMS has to implement
some form of a lease system to invalidate readers for transactional workloads that have
stricter latency requirements. In addition to introducing complexity in the concurrency
control protocol of the DBMS, this approach also requires that the client knows beforehand
the address of the blocks it needs to access, which requires a separate RPC service or
some external directory maintained by the DBMS to convey this information. We envision
these challenges to be non-trivial in achieving server-side RDMA.
\\ \vspace{-0.1in}

%% -----------------------
%% Shipping Compute
%% -----------------------
\textbf{Shipping Computation to Data:} 
Server- and client-side RDMA allow external tools to access data with extremely low data
export overhead. The problem, however, is that RDMA requires specialized hardware and is
only viable when the application is co-located in the same data center as the DBMS. The
latter is unlikely for a data scientist working on a personal workstation. A deeper issue
is that using RDMA requires the DBMS to ``pull'' data to the computational resources that
execute the query. The limitations of this approach are widely known, especially in the
case where server-side filtering is difficult to achieve. %  (e.g., server-side RDMA). 

If we adopt a ``push'' the query to the data approach, then using native Arrow storage in
the DBMS does not provide benefits to network speed. Instead, we can leverage Arrow as an
API between the DBMS and external tools to improve programmability. Because Arrow is a
standardized memory representation of data, if external tools support Arrow as input, then
it is possible to run the program on the DBMS by replacing Arrow references with mapped
memory images from the DBMS process. Using shared memory in this manner introduces a new
set of problems involving security, resource allocation, and software engineering. By
making an analytical job portable across machines, it also allows dynamic migration of a
task to a different server. In combination with RDMA, this leads to true serverless HTAP
processing where the client specifies a set of tasks, and the DBMS dynamically assembles a
heterogeneous pipeline with low data movement cost.

%% ==================================================================
%% Evaluation
%% ==================================================================
\section{Evaluation}
\label{sec:eval}
We next present an experimental analysis of our system. We implemented our storage engine
in the \sysname DBMS~\cite{cmu-dbms}. We performed our evaluation on a machine with a
dual-socket 10-core Intel Xeon E5-2630v4 CPU, 128~GB of DRAM, and a 500~GB Samsung 970 EVO
Plus SSD. For each experiment, we use \texttt{numactl} to interleave memory allocation on
available NUMA regions. All transactions execute as JIT-compiled stored procedures with
logging enabled. We run each experiment ten times and report the average.

We first evaluate our OLTP performance and quantify performance interference from the
transformation process. We then provide a set of micro-benchmarks to study the performance
characteristics of the transformation algorithm. Finally, we compare data export
performance in our system against current approaches. 

%% ---------------------------------------------------------------
%% OLTP Performance
%% ---------------------------------------------------------------
\subsection{OLTP Performance}
\label{sec:eval-oltp}
We measure the DBMS's OLTP performance to demonstrate the viability of
our storage architecture and that our transformation process is lightweight. 
We use TPC-C~\cite{tpc-c} in this experiment with one warehouse per client. \sysname uses
the OpenBw-Tree for all indexes~\cite{wang18}. All transactions are submitted open-loop.
We report the DBMS's throughput and the state of blocks at the end of each run. We use
\texttt{taskset} to limit the number of available CPU cores as the sum of worker, logging, 
and GC threads. To account for the additional resources 
required by transformation, the system has one logging thread, one transformation thread,
and one GC thread for every 8 worker threads. We deploy the DBMS with three transformation
configurations: (1) disabled, (2) variable-length gather, and (3) dictionary compression.
For trials with \sysname's block transformation enabled, we use an aggressive threshold
time of 10~ms and only target the tables that generate cold data: \dbTable{order},
\dbTable{order\_line}, \dbTable{history}, and \dbTable{item}. In each run, the compactor
attempts to process all blocks from the same table in the same group.

\begin{figure}[t!]
    \centering
    \subfloat[Throughput]{
        \includegraphics[width=\columnwidth]{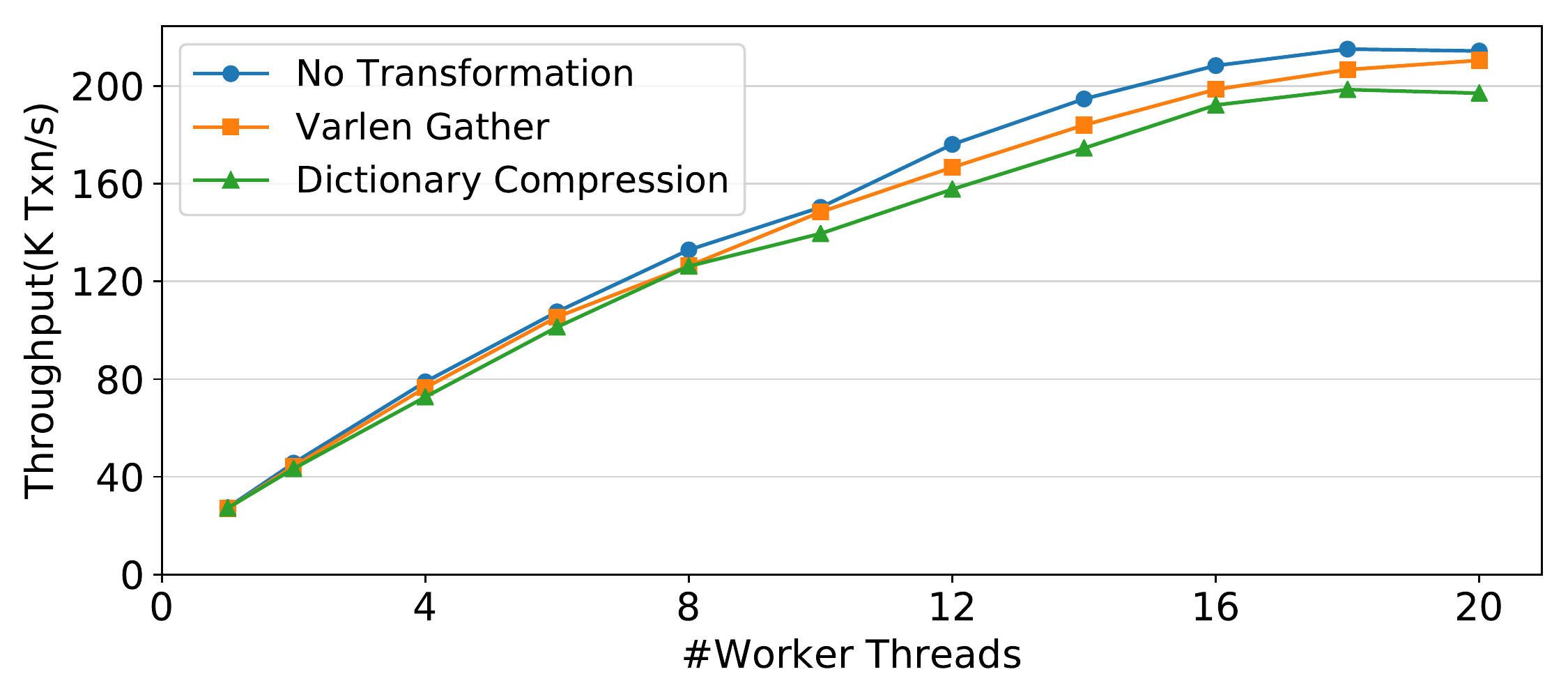}
        \label{fig:oltp-throughput}
    }
    \\
    \subfloat[Block State Coverage]{
        \includegraphics[width=\columnwidth]{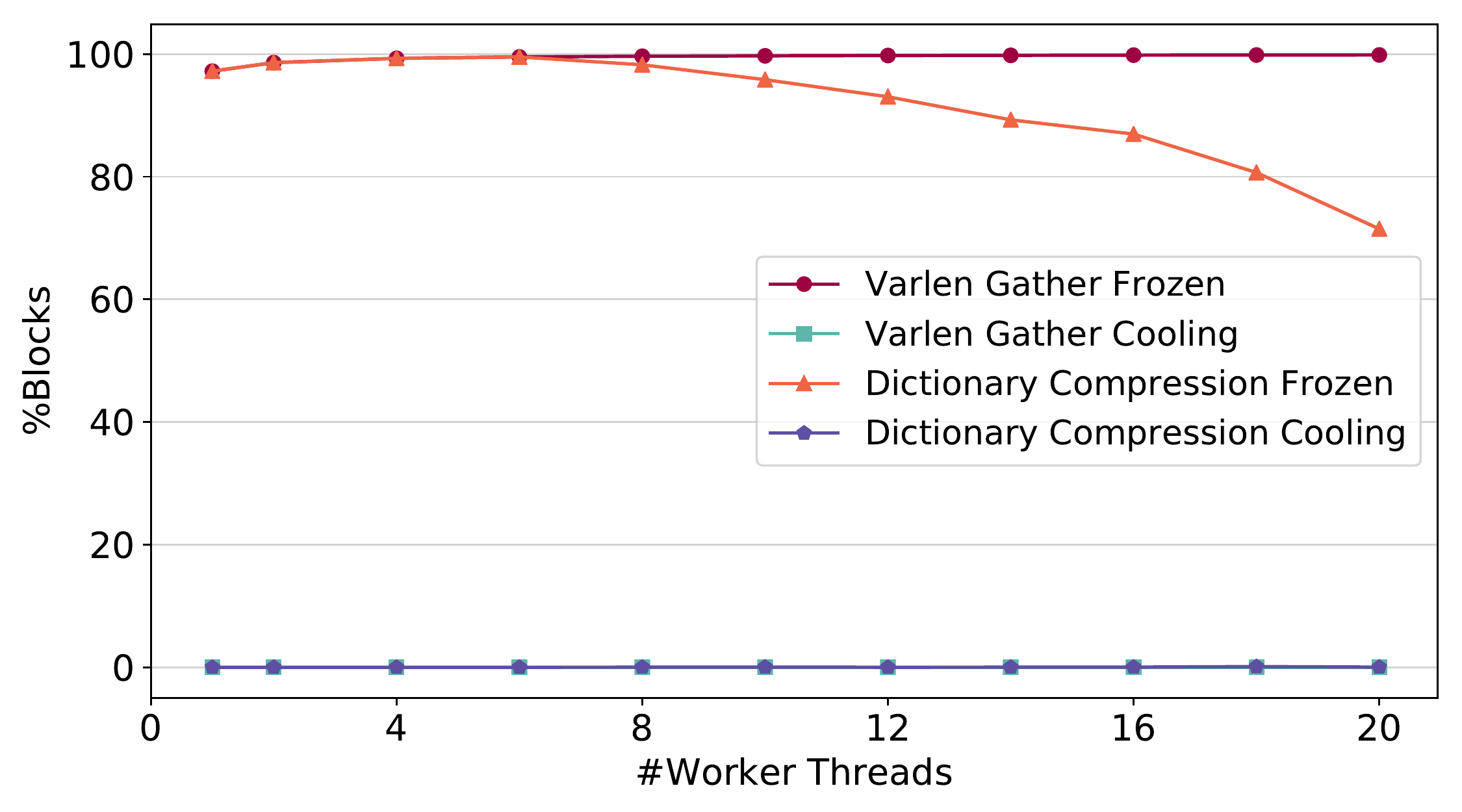}
        \label{fig:oltp-blockstate}
    }
    \caption{
        \textbf{OLTP Performance} -- 
        Runtime measurements of the DBMS for the TPC-C workload, varying the number of
        threads.
    }
    \label{fig:oltp}
\end{figure}

The results in \cref{fig:oltp-throughput} show that the DBMS achieves good scalability
and incurs little overhead from the transformation process (at most 10\%). The
interference is more prominent as the number of workers increases due to more work for
the transformation thread. At 20 worker threads, the DBMS's scaling degrades.
This decrease is because our machine only has 20 physical CPU cores, and threads no longer
have dedicated cores. The problem of threads swapping is worse with the additional
transformation thread. Dictionary compression has a larger performance impact because it
is computationally more intensive.

% To better understand this issue, we measured the abort rate of transactions and the
% \# of transactions that was stalled due to the transformation process. We did not
% observe a statistically significant amount of change in abort rates and a negligible
% \# of stalled transactions (<0.01\%).

In \cref{fig:oltp-blockstate}, we report the percentage of blocks in the \statusCooling
and \statusFrozen state at the end of each run. We omit results for the \dbTable{item}
table because it is a read-only table, and its blocks are always \statusFrozen.
%Because the DBMS does not transform any blocks that it can insert into, these blocks remain
% in the \statusHot state, and thus, we do not achieve 100\% block coverage.
These results show that the DBMS achieves nearly complete coverage, but starts to lag for
a higher number of worker threads in the case of dictionary compression. This is because 
dictionary compression is an order of magnitude slower than simple gathering, as we will
show in \cref{sec:eval-arrow}. Per our design goal, the transformation process yields
resources to user transactions in this situation and does not result in a significant drop
in transactional throughput. As discussed in \cref{sec:transformation-additional}, one can
simply parallelize the transformation process by partitioning based on block address when
the transformation thread is lagging behind. To achieve full transformation, we ran the 
benchmark with one additional transformation thread, and observe an additional \~15\%
reduction in transactional throughput. 
% In a real workload, it is unlikely that the DBMS sustains high
% load continuously, and the transformation process will eventually catch up in
% transformation rate when system load drops.
% One could manually control resouce allocation between transactions and the transformation thread, 
% instead of leaving control to the OS scheduler, to avoid over-saturating the
% transformation process. 
\\ \vspace{-0.1in}

%% -----------------------
%% Row vs. Column
%% -----------------------
\textbf{Row vs. Column:}
To investigate the impact of using a column-store with an OLTP workload, we run a 
synthetic micro-benchmark that compares our storage architecture against a row-store. We
simulate a row-store by declaring a single, large column that stores all of a tuples'
attributes contiguously. Each attribute is an 8-byte fixed-length integer. We fix the number
of threads executing queries and scale up the number of attributes per tuple from one to
64. We run a workload comprised of either (1) insert or (2) update queries
(10 million each) and report the throughput. We ignore the overhead of maintaining indexes
in our measurements as this cost is the same for both storage models.

\begin{figure}[t!]
    \centering
    \includegraphics[width=\columnwidth]{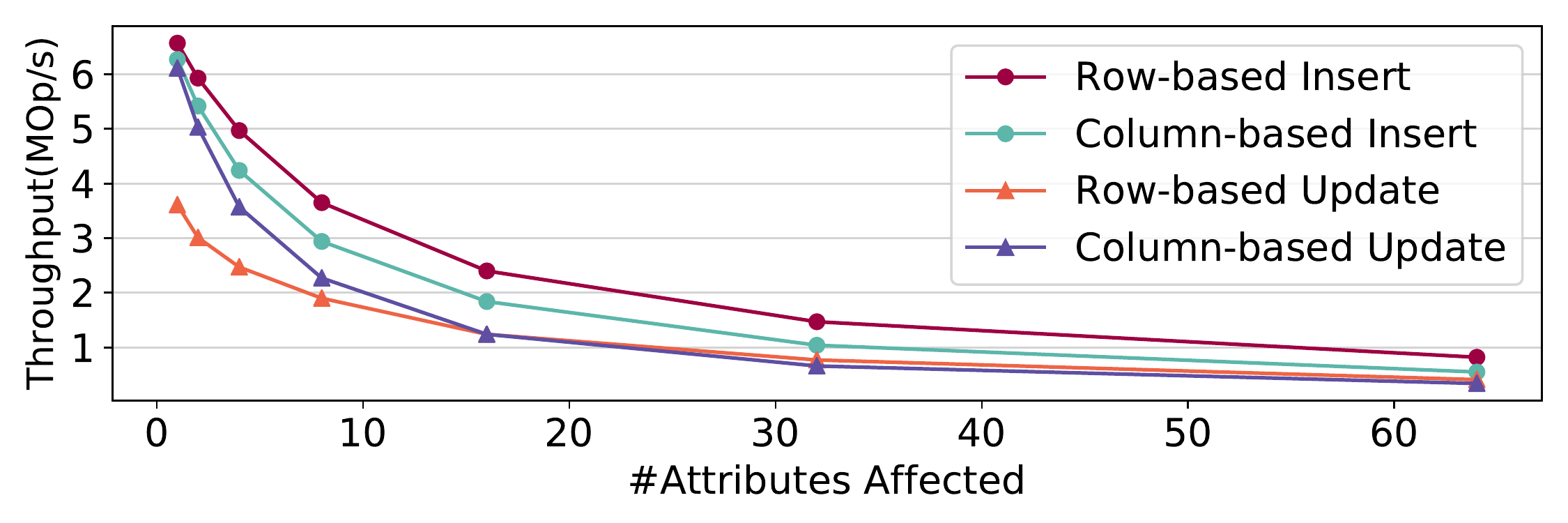}
    \caption{
        \textbf{Row vs. Column} -- 
        Measurements of raw storage speed of \sysname,
        row vs. column, varying number of attributes modified.
        For inserts, the x-axis is the number of attributes of the inserted tuple; for 
        updates, it is the number of attributes updated. 
    }
    \label{fig:row-v-col}
\end{figure}

The results in \cref{fig:row-v-col} show that the two approaches do not exhibit a large
performance difference. Even for the insert workload, where raw memory copy speed matters
more, the gap never exceeds 40\%. For the update workload, a column-store outperforms row
stores when the number of attributes copied is small due to a smaller memory footprint. As
the number of attributes grows, the row-store becomes slightly faster than the 
column-store, but with a much lower difference due to fixed-cost of maintaining versions. 
These results indicate that it is unlikely that an optimized row-store will provide a 
compelling performance improvement in an in-memory setting over a column-store.
%In a full system, a
%query incurs even more overhead in index maintenance and other
%internal tasks. It is unlikely that a row-store can deliver significant speed-up in an
% in-memory setting. 

%% ---------------------------------------------------------------
%% Transformation to Arrow
%% ---------------------------------------------------------------
\subsection{Transformation to Arrow}
\label{sec:eval-arrow}

\begin{figure*}[t!]
    \centering
    \fbox{
        \includegraphics[width=\columnwidth]{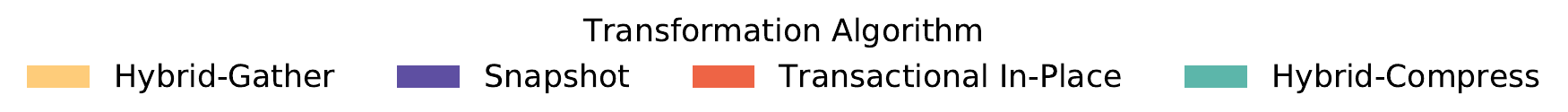}
    }
    \\
    \subfloat[Throughput (50\% Variable-Length Columns)]{
        \includegraphics[width=\columnwidth]{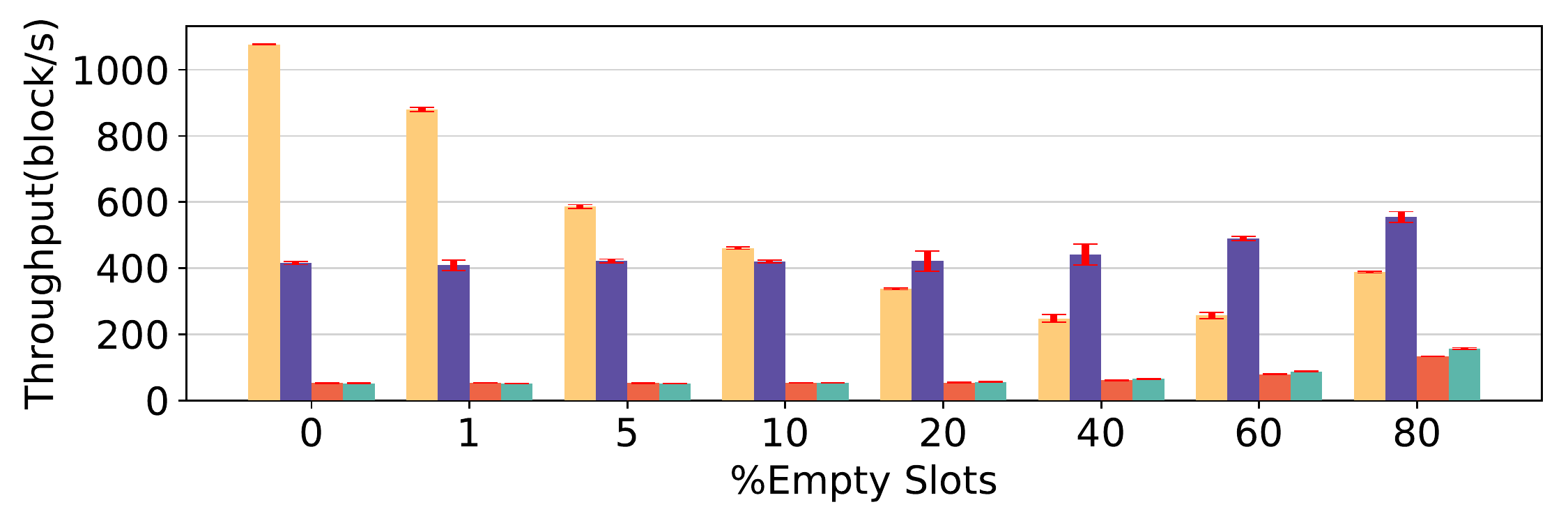}
        \label{fig:transform-throughput}
    }
    \hfill
    \subfloat[Performance Breakdown (50\% Variable-Length Columns)]{
        \includegraphics[width=\columnwidth]{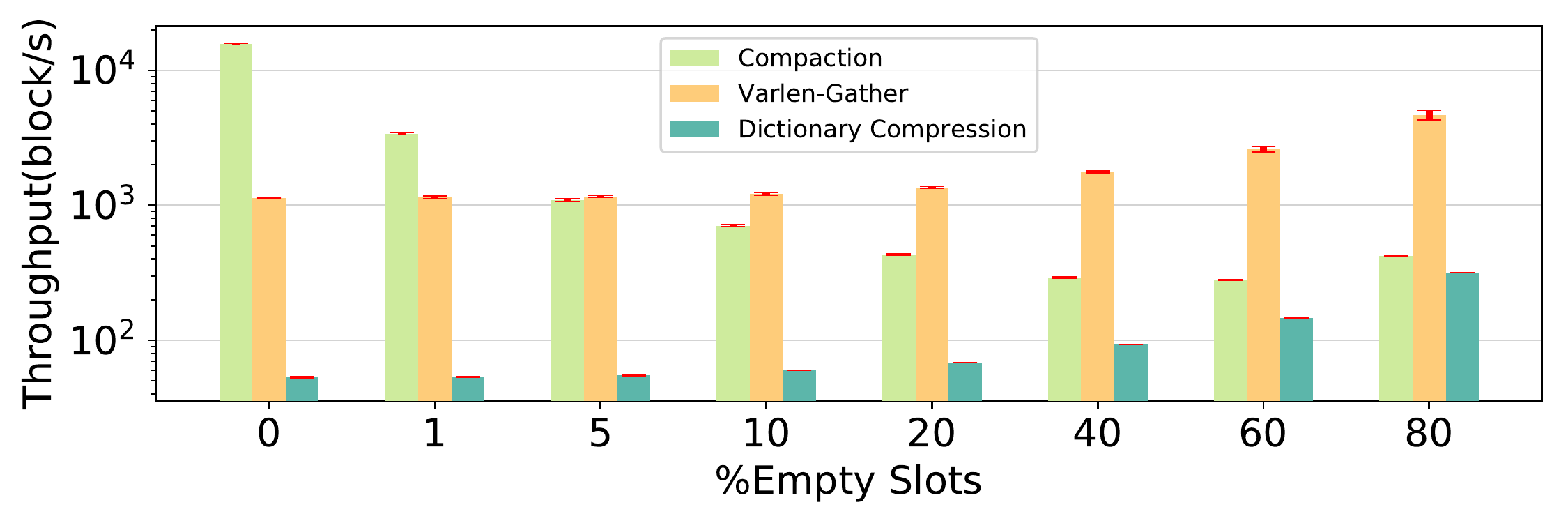}
        \label{fig:transform-breakdown}
    }
    \\
    \subfloat[Throughput (Fixed-Length Columns)]{
        \includegraphics[width=\columnwidth]{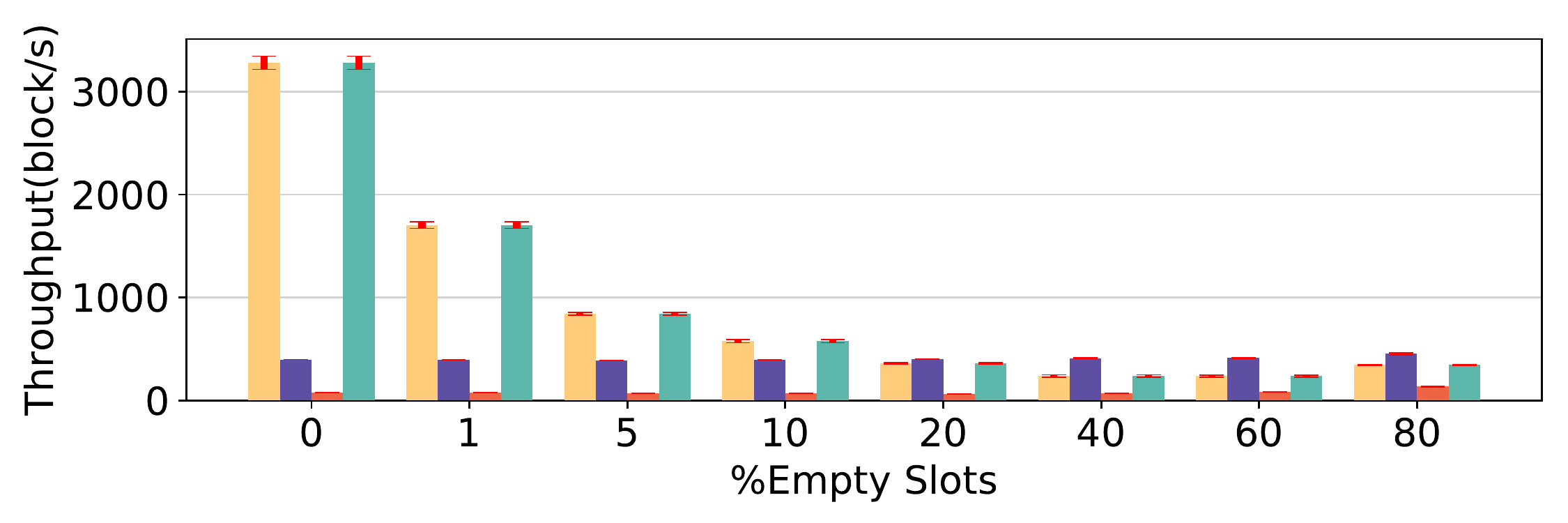}
        \label{fig:transform-throughput-fixed}
    }
    \hfill
    \subfloat[Throughput (Variable-Length Columns)]{
        \includegraphics[width=\columnwidth]{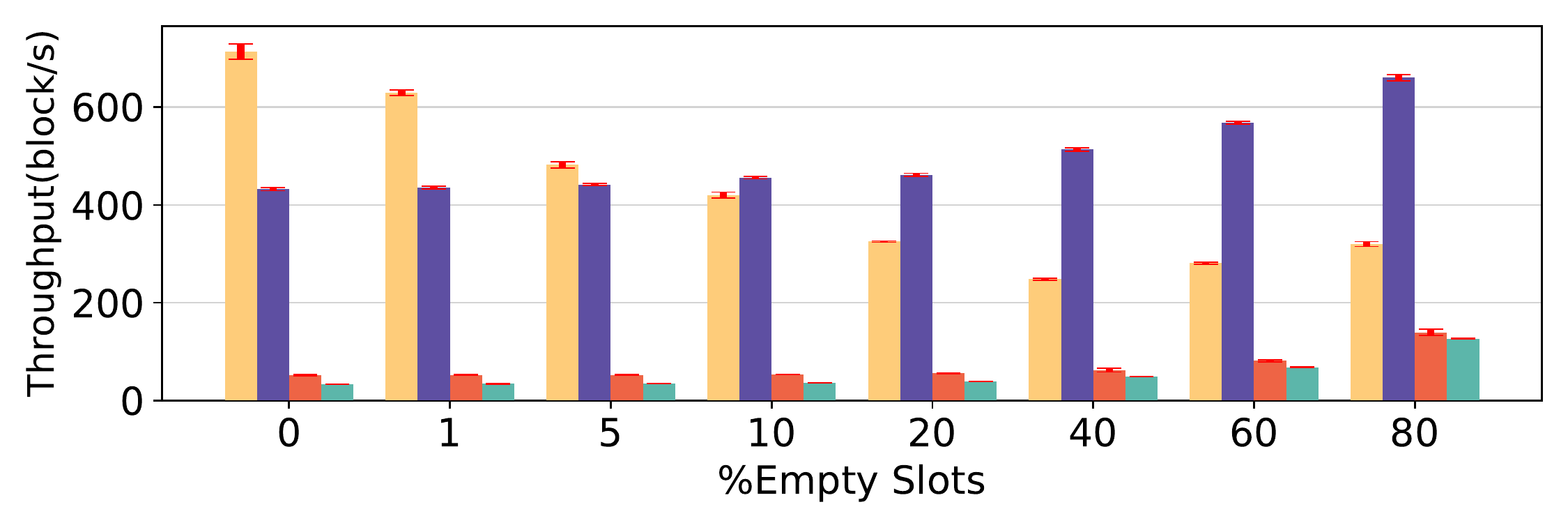}
        \label{fig:transform-throughput-varlen}
    }
    \caption{
        \textbf{Transformation Throughput} -- 
        Measurements of the DBMS's transformation algorithm throughput and movement cost
        when migrating blocks from the relaxed format to the canonical Arrow format.
    }
    \label{fig:transform}
\end{figure*}

We next evaluate our transformation algorithm and analyze the effectiveness of each
sub-component. We use micro-benchmarks to demonstrate the DBMS's performance when
migrating blocks from the relaxed Arrow format to their canonical form. Each trial of
this experiment simulates one transformation pass in a system to process data that has
become cold since the last invocation. 

The database used has a single table of $\sim$16M tuples with two columns: (1) a 8-byte
fixed-length column and (2) a variable-length column with values between 12--24 bytes.
Under this layout, each block holds $\sim$32K tuples.  We also ran these same experiments
on a table with more columns or larger varlens,  but did not observe a major difference in
trends. An initial transaction populates the table, and inserts empty tuples at random
to simulate deletion.
\\ \vspace{-0.1in}

%% -----------------------
%% Throughput
%% -----------------------
\textbf{Throughput:}
Recall from \cref{sec:transformation-algo} that our transformation algorithm is a hybrid
two-phase implementation. For this experiment, we assume there is no concurrent
transactions and run the two phases consecutively without waiting. We benchmark both
versions of our algorithm: (1) gathering variable-length values and copying them into a 
contiguous buffer (\transformHybridGather) and (2) using dictionary compression on
variable-length values (\transformHybridCompress). We also implemented two baseline
approaches for comparison purposes: (1) read a snapshot of the block in a transaction 
and copy into a Arrow buffer using the Arrow API (\transformSnapshot) and (2) 
perform the entire transformation in-place in a transaction (\transformTxnInplace).
We use each algorithm to process 500 blocks (1~MB each) and vary the percentage of empty
slots in each run.

\begin{figure}[t!]
    \centering
    \includegraphics[width=\columnwidth]{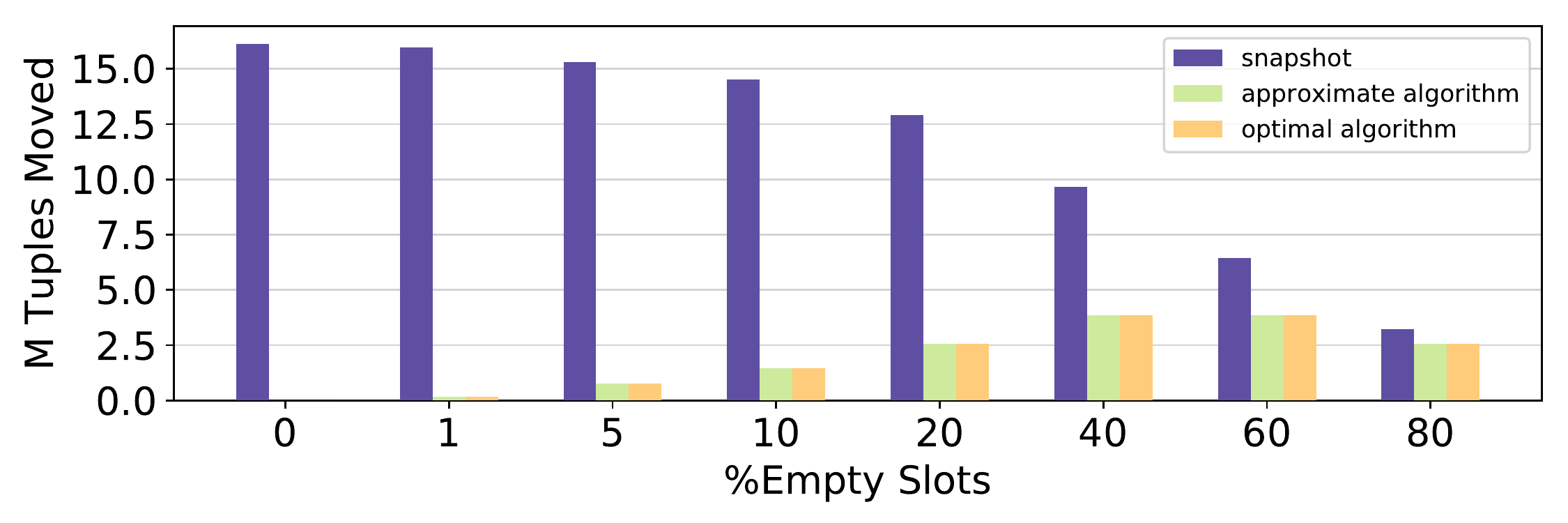}
    \caption{
        \textbf{Write Amplification} --
        Total write amplification is number of tuple movement times a constant for each
        table, determined by the layout and number of indexes on that table.
    }
    \label{fig:movement}
\end{figure}

\begin{figure*}[t!]
    \centering
    \fbox{
        \includegraphics[width=\columnwidth]{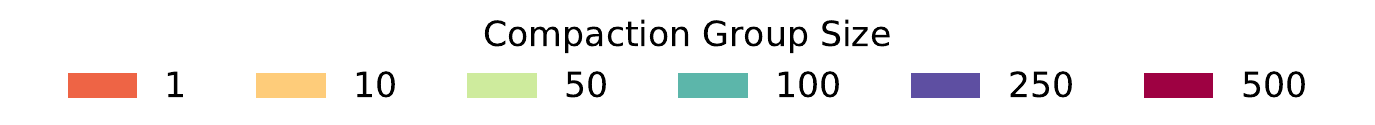}
    }
    \\
    \subfloat[Number of Blocks Freed]{
        \includegraphics[width=\columnwidth]{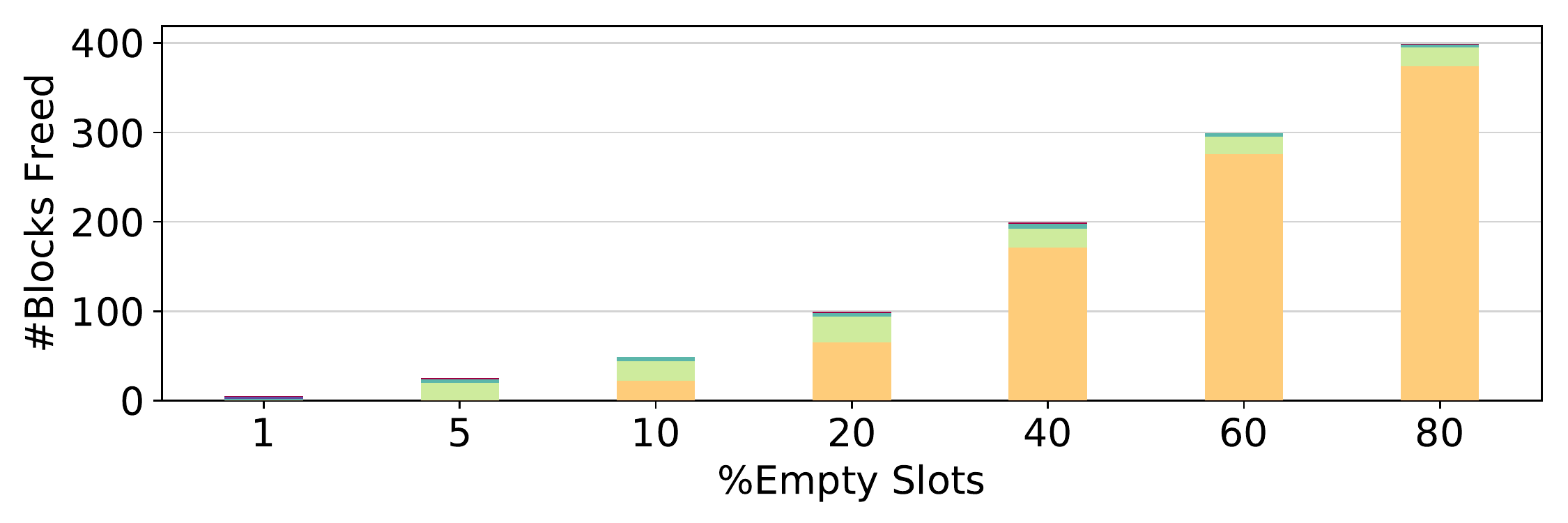}
        \label{fig:groupsize-freed}
    }
    \hfill
    \subfloat[Write-Set Size of Transactions]{
        \includegraphics[width=\columnwidth]{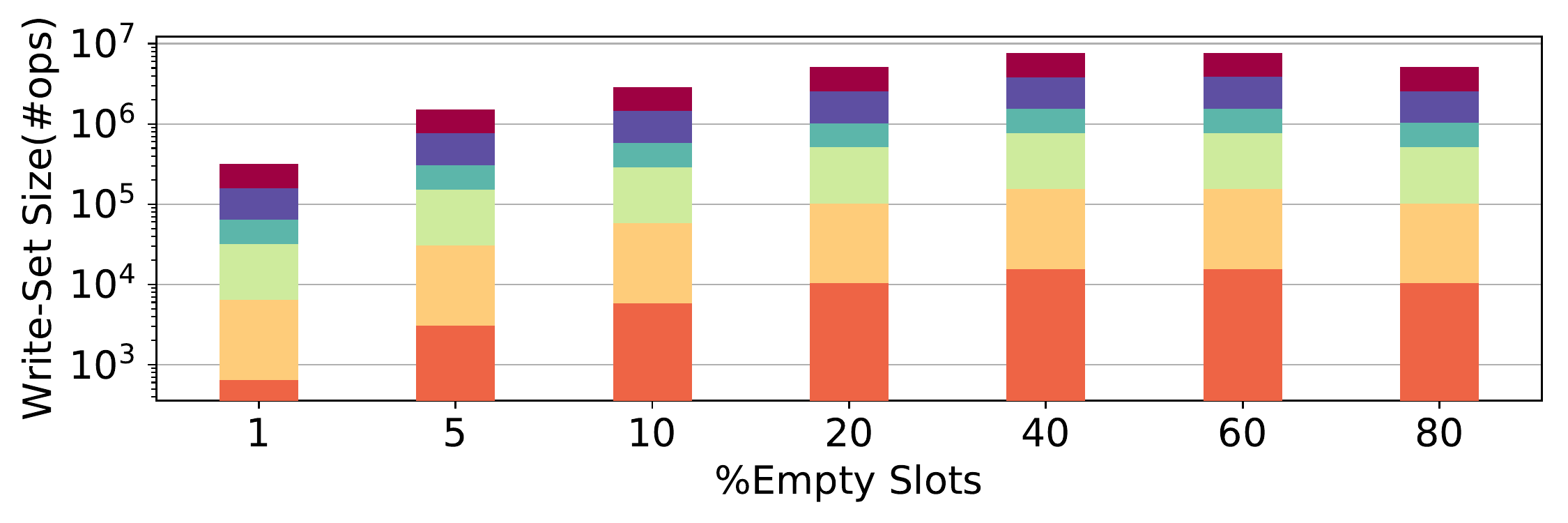}
        \label{fig:groupsize-writeset}
    }
    \caption{
        \textbf{Sensitivity on Compaction Group Size} --
        Efficacy measurements of the transformation algorithm when varying the number of 
        blocks per compaction group while processing 500 blocks. The percentage of empty 
        slots is what portion of each block is empty (i.e., does not contain a tuple).
        (\ref{fig:groupsize-freed}) shows the number of blocks freed during one round.
        (\ref{fig:groupsize-writeset}) shows the number of operations processed per second.
    }
    \label{fig:groupsize}
\end{figure*}

\begin{figure}[t!]
    \centering
    \includegraphics[width=\columnwidth]{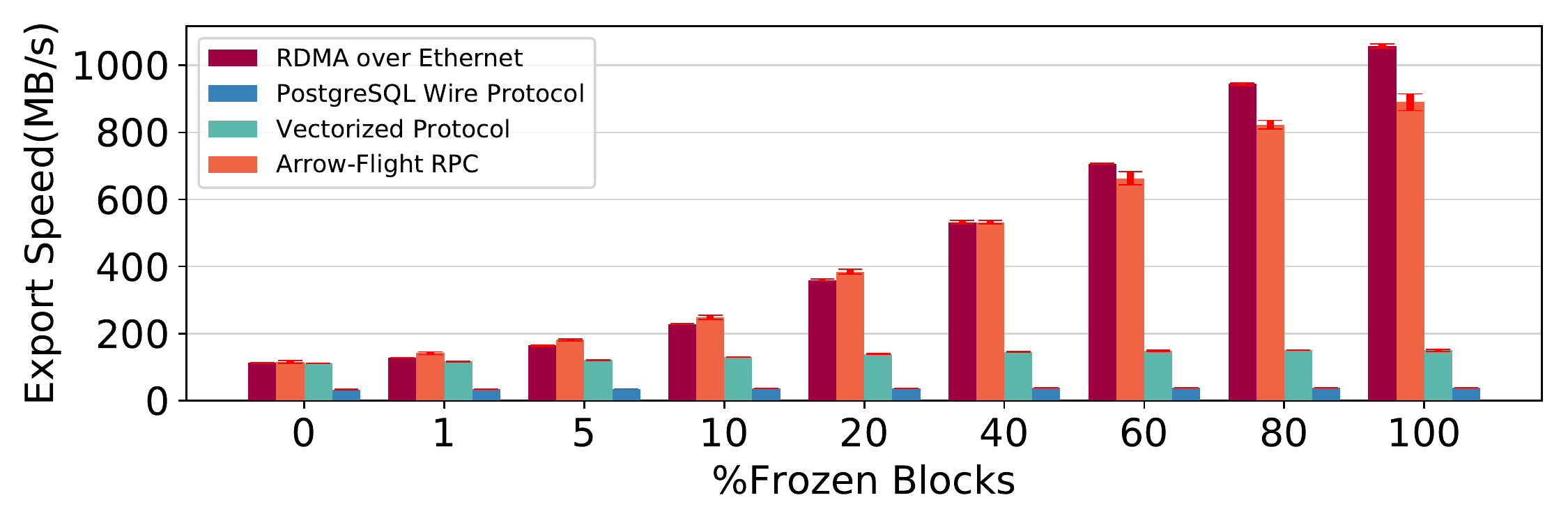}
    \caption{
        \textbf{Data Export} -- Measurements of export speed with different export
                                mechanisms in \sysname, varying \% of \statusHot blocks. 
    }
    \label{fig:export}
\end{figure}

The results in \cref{fig:transform-throughput} show that \transformHybridGather
outperforms the alternatives, achieving sub-millisecond performance when blocks are mostly
full (i.e., the number of empty slots is less than 5\%). Performance drops as emptiness
increases since the DBMS needs to move more tuples. Such movement is an order of magnitude
more expensive than \transformSnapshot due to the random memory access pattern. As the
blocks become more than half empty, the number of tuples that the DBMS moves decreases,
and thus the throughput bounces back. \transformTxnInplace performs poorly because of the
version maintenance overhead. \transformHybridCompress is also an order of magnitude
slower than \transformHybridGather and \transformSnapshot because building the dictionary 
is computationally expensive.

To understand the behavior of these algorithms better, we provide a breakdown of each
phase in \cref{fig:transform-breakdown}. We present the graph in log-scale due to the
large range of performance changes. When the number of empty slots in a block is low
(i.e., <5\%), the DBMS completes the compaction phase in microseconds because it is
reduced to a bitmap scan. In this best-case scenario, the cost of variable-length gather
dominates. The performance of the compaction phase drops as the number of empty slots
increases and starts to dominate the cost of \transformHybridGather at 5\% empty.
Dictionary compression is always the bottleneck in \transformHybridCompress.

We next measure how the column types affect the performance of the four transformation
algorithms. We run the same micro-benchmark but make the database's columns either all
fixed-length (\cref{fig:transform-throughput-fixed}) or variable-length
(\cref{fig:transform-throughput-varlen}). These results show that the general performance
trend does not change based on the data layouts. \transformSnapshot performs better when
there are more variable-length values in a block because it does not update nor copy the
metadata associated with each value. Given this, we show only 50\% variable-length columns
results for other experiments.
\\ \vspace{-0.1in}

%% -----------------------
%% Write Amplification
%% -----------------------
\textbf{Write Amplification:}
The previous throughput results show that \transformSnapshot outperforms our hybrid
algorithm when blocks are $\sim$20\% empty. These measurements, however, fail to capture
the overhead of updating the index entries for any tuples that change their physical
location in memory~\cite{wu17}. The effect of this write amplification depends on the type
and number of indexes on the table, but the cost for each tuple movement is constant.
Therefore, it suffices to measure the total number of tuple movements that trigger index
updates. The \transformSnapshot algorithm always moves every tuple in the compacted
blocks. We compare its performance against the approximate and optimal algorithms from
\cref{sec:transformation-algo}.

As shown in \cref{fig:movement}, our algorithm is several orders of magnitudes more
efficient than \transformSnapshot in the best case, and twice as efficient when the blocks
are half empty. The gap narrows as the number of empty slots per block increases. There is
little difference in the result of the approximate algorithm versus the optimal algorithm,
validating our decision to use the approximate algorithm to save one pass through the
blocks during transformation.
%That is, the approximate approach generates almost the same
%physical configuration for blocks as the optimal approach.
%Given this, and that the optimal algorithm requires one more scan across the blocks
%than the approximate one, we conclude that the marginal improvement does not justify
%the cost. Thus, we use the approximate algorithm for all other experiments.
\\ \vspace{-0.1in}

%% -----------------------
%% Compaction Group Size
%% -----------------------
\textbf{Sensitivity on Compaction Group Size:}
For the next experiment, we evaluate the effect of the compaction group size on
performance. The DBMS groups blocks together for compaction and then frees any empty
blocks. This grouping enables the DBMS to reclaim memory from deleted slots. The size of
each compaction group is a tunable parameter in the system. Larger group sizes result in
the DBMS freeing more blocks but increases the size of the write-set for compacting
transactions, which increases the likelihood that they will abort due to a conflict. We
use the same setup from the previous experiment, performing a single transformation pass
through 500 blocks while varying group sizes. % to evaluate the trade-offs.

\cref{fig:groupsize-freed} shows the number of freed blocks with different compaction
group sizes. When blocks are only 1\% empty, larger group sizes are required to release
any memory. As the vacancy rate of blocks increases, smaller group sizes perform
increasingly well, and larger values bring only marginal benefit. We show the cost of
larger transactions as the size of their write-sets in \cref{fig:groupsize-writeset}.
These results indicate that larger group sizes increase transactions' write-set size, but
yield a diminishing return on the number of blocks freed. The ideal fixed group size is
between 10 and 50, which balances good memory reclamation and relatively small write-sets.
To achieve the best possible performance, the DBMS should employ an intelligent policy
that dynamically forms groups of different sizes based on the blocks it is compacting. We 
defer this problem as future work.
% Overall, our solution is orders of magnitude faster than the na\"{\i}ve baseline in the
% common case where the blcoks are mostly full, and achieves comparable performance when
% considering write amplification when blocks are mostly empty.

%% ---------------------------------------------------------------
%% Data Export
%% ---------------------------------------------------------------
\subsection{Data Export}
\label{sec:eval-export}
For this last experiment, we evaluate the DBMS's ability to export data to an external
tool. We compare four of the data export methods from \cref{sec:export}: (1) client-side
RDMA, (2) the Arrow Flight RPC, (3) vectorized wire protocol from \cite{raasveldt17}, and
(4) row-based \postgres wire protocol. We implement (3) and (4) in \sysname according to
their specifications. Because RDMA requires specialized hardware, we run these experiments
on two different servers with eight-core Intel Xeon D-1548 CPUs, 64~GB of DRAM, and a
dual-port Mellanox ConnectX-3 10~GB NIC (PCIe v3.0, eight lanes).

We use the TPC-C \dbTable{order\_line} table with 6000 blocks ($\sim$7~GB total size, 
including variable-length values) on the server. On the client-side, we run a Python 
application, and report the time taken between sending a request for data and the
beginning of execution of analysis. For each export method, we write a corresponding
client-side protocol in C++, and use Arrow's cross-language API~\cite{pyarrow-api} to
expose it to the Python program in a zero-copy fashion. The client runs a TensorFlow
program that passes all data through a single linear unit, as the performance of this
component is irrelevant to our system. We vary the percentage of blocks frozen in the
DBMS to study the effect of concurrent transactions on export speed. Recall that if a
block is not frozen, the DBMS must materialize it transactionally before sending.

The results in \cref{fig:export} shows that \sysname exports data orders-of-magnitude
faster than the base-line implementations. When all blocks are \statusFrozen, RDMA
saturates the available network bandwidth, and Arrow Flight can utilize up to 80\% of the
available network bandwidth. When the system has to materialize every block, the
performance of Arrow Flight drops to be equivalent to the vectorized wire protocol. RDMA
performs slightly worse than Arrow Flight with a large number of hot blocks, because
Flight has the materialized block in its CPU cache, whereas the NIC bypasses this cache
when sending data. Both the \postgres wire protocol and the vectorized protocol do not
benefit much from eliding transactions on cold, read-only data. Hence, this experiment
indicates that the main bottleneck of the data export process in a DBMS is this
serialization/deserialization step. Using Arrow as a drop-in replacement wire protocol in
the current architecture does not achieve its full potential. Instead, storing data in a
common format reduces this cost and boosts data export performance.

%% ==================================================================
%% Related Work
%% ==================================================================
\section{Related Work}
\label{sec:relatedwork}
We presented our system for high transaction throughput on a storage format optimized for
analytics, and now discuss three key facets of related work. In particular, we provide an
overview of other universal storage formats, OLTP systems on column-stores, and
optimizations for DBMS data export.
\\ \vspace{-0.1in}

%% -----------------------
%% Universal Storage Formats
%% -----------------------
\textbf{Universal Storage Formats:}
The idea of building a data processing system on top of universal storage formats has been 
explored in other implementations. Systems such as Apache Hive~\cite{hive}, Apache 
Impala~\cite{impala}, Dremio~\cite{dremio}, and OmniSci~\cite{omnisci} support data
ingestion from universal storage formats to lower the data transformation cost. These are
analytical systems that ingest data already generated in the format from an OLTP system,
whereas our DBMS natively generates data in the storage format as a data source for these
systems. 

Among the storage formats other than Arrow, Apache ORC~\cite{apacheorc} is the most
similar to our DBMS in its support for ACID transactions. ORC is a self-describing
type-aware columnar file format designed for Hadoop. It divides data into \textit{stripes}
that are similar to our concept of blocks. Related to ORC is Databricks' Delta Lake
engine~\cite{deltalake} that acts as a ACID transactional engine on top of cloud storage.
These solutions are different from our system because they are intended for incremental
maintenance of read-only data sets and not high-throughput OLTP. Transactions in these
systems are infrequent, not performance critical, and have large write-sets. Apache
Kudu~\cite{kudu} is an analytical system that is similar in architecture to our system,
and integrates natively with the Hadoop ecosystem. However, transactional semantics in
Kudu is restricted to single-table updates or multi-table scans and does not support
general-purpose SQL transactions~\cite{kudu-txn}.
\\ \vspace{-0.1in}
%\todo{write about HDF}

%% -----------------------
%% Compaction Group Size
%% -----------------------
\textbf{OLTP on Column-Stores:}
Since \citeauthor{pax} first introduced the PAX model~\cite{pax}, the community has
implemented several systems that supports transactional workloads on column-stores. PAX
stores data in columnar format, but keeps all attributes of a single tuple within a disk
page to reduce I/O cost for single tuple accesses. HYRISE~\cite{hyrise} improved upon this
scheme by vertically partitioning each table based on access patterns. SAP
HANA~\cite{sikka12} implemented migration from row-store to column-store in addition to
partitioning. MemSQL's SingleStore~\cite{memsql} improved their transactional performance
on columnar data by adding hash indexes, sub-segment access, and fine-grain locking. Writes
are absorbed by an in-memory skip list, while deletes are marked directly in the columnar
data. Background optimization routines are responsible for eventually flushing and
compacting the results of these operations. Peloton~\cite{arulraj16} introduced
the logical tile abstraction to enable migration without a need for disparate execution
engines. Our system is most similar to HyPer~\cite{lang, kemper11, hyper12, neumann15}
and L-Store~\cite{lstore}. HyPer runs exclusively on columnar format and guarantees ACID
properties through a multi-versioned delta-based concurrency control mechanism similar to
our system; it also implements a compression for cold data chunks by instrumenting the OS
for access observation. Our system is different from HyPer in that it is built around the
open-source Arrow format and provides native access to it. HyPer's hot-cold transformation
also assumes heavy-weight compression operations, whereas our transformation process is
designed to be fast and computationally inexpensive, allowing more fluid changes in a
block's state. L-Store also leverages the hot-cold separation of tuple access to allow
updates to be written to \textit{tail-pages} instead of more expensive cold storage. In
contrast to our system, L-Store achieves this through tracing data lineage and an
append-only storage within the table itself. 
\\ \vspace{-0.1in}

%% -----------------------
%% Networking for DBMS
%% -----------------------
\textbf{Optimized DBMS Networking:}
There has been considerable work on using RDMA to speed up DBMS workloads. IBM's DB2 
pureScale~\cite{purescale} and Oracle Real Application Cluster (RAC)~\cite{rac} use
RDMA to exchange database pages and achieve shared-storage between nodes. Microsoft
Analytics Platform Systems~\cite{msaps} and Microsoft SQL Server with SMB
Direct~\cite{smbdirect} utilize RDMA to bring data from a separate storage layer to
the execution layer. \citeauthor{binnig16}~\cite{binnig16} and 
\citeauthor{farm15}~\cite{farm15} proposed using RDMA for distributed transaction 
processing~\cite{binnig16}. Li et al.~\cite{li16} proposed
a method for using RDMA to speed up analytics with remote memory. All of these work
attempts to improve the performance of distributed DBMS through using RDMA within the
cluster. Our paper looks to improve efficiency across the data processing pipeline through
better interoperability with external tools.

\citeauthor{raasveldt17} demonstrated that transferring large
amounts of data from the DBMS to a client is expensive over existing wire row-oriented
protocols (e.g., JDBC/ODBC)~\cite{raasveldt17}. They then explored how to improve server-side 
result set
serialization to increase transmission performance. A similar technique was proposed in
the olap4j extension for JDBC in the early 2000s~\cite{olap4j}. These works, however,
optimize the DBMS's network layer, whereas this paper tackles the challenge more broadly
through changes in both the network layer and the underlying DBMS storage.

%% ==================================================================
%% CONCLUSION
%% ==================================================================
\section{Conclusion}
\label{sec:conclusion}
We presented \sysname's Arrow-native storage architecture for in-memory OLTP workloads.
The system implements a multi-versioned, delta-store transactional engine capable of 
directly emitting Arrow data to external analytical tools. To ensure OLTP performance,
the system allows transactions to work with a relaxed Arrow format and employs a
lightweight in-memory transformation process to convert cold data into full Arrow in
milliseconds. This allows the DBMS to support bulk data export to external analytical
tools at zero serialization overhead. We evaluated our implementation, and show good OLTP
performance while achieving orders-of-magnitudes faster data export compared to current
approaches. 

% TODO: Add this back in for the camera ready version
\noindent \textbf{Acknowledgements.}
This work was supported (in part) by funding from the U.S.
by the National Science Foundation under awards
\href{https://www.nsf.gov/awardsearch/showAward?AWD_ID=1846158}{IIS-1846158},
\href{https://www.nsf.gov/awardsearch/showAward?AWD_ID=1718582}{IIS-1718582},
and \href{https://www.nsf.gov/awardsearch/showAward?AWD_ID=1822933}{SPX-1822933},
Google Research Grants, and the
\href{https://sloan.org/grant-detail/8638}{Alfred P. Sloan Research Fellowship} program.

%% ==================================================================
%% BIBLIOGRAPHY
%% ==================================================================
\balance

\bibliographystyle{ACM-Reference-Format}
\bibliography{storage-arrow}

\end{document}